\documentclass[aps,twocolumn,prb,amsmath,amssymb,superscriptaddress]{revtex4-1}
\usepackage{url}
\usepackage{graphicx}
\usepackage{placeins}
\usepackage{color}
\usepackage{natbib}
\usepackage{hyperref}
\usepackage{notes2bib}
\usepackage{mathbbol}
\usepackage{float}
\usepackage{ulem}
\usepackage{tikz}
\usepackage{xcolor}
\usepackage{svg}
\usepackage{caption}
\usepackage{dsfont}

\usepackage{comment}
\usepackage{amsfonts}
\usepackage{graphicx}
\usepackage{makecell}

\newcommand{\bpm}{\begin{pmatrix}}
\newcommand{\epm}{\end{pmatrix}}

\newcommand{\be}{\begin{equation}}
\newcommand{\ee}{\end{equation}}
\newcommand{\beq}{\begin{eqnarray}}
\newcommand{\eeq}{\end{eqnarray}}

\newcommand{\up}{\uparrow}
\newcommand{\down}{\downarrow}

\begin{document}
\title{Superconductivity in monolayer and few-layer graphene:\\ III Impurity-induced subgap states and quasi-particle interference patterns}

\author{Emile Pangburn}
\affiliation{Institut de Physique Th\'eorique, Universit\'e Paris Saclay, CEA
CNRS, Orme des Merisiers, 91190 Gif-sur-Yvette Cedex, France}
\author{Louis Haurie}
\affiliation{Institut de Physique Th\'eorique, Universit\'e Paris Saclay, CEA
CNRS, Orme des Merisiers, 91190 Gif-sur-Yvette Cedex, France}
\author{Adeline Cr\'epieux}
\affiliation{Aix Marseille Univ, Universit\'e de Toulon, CNRS, CPT, Marseille, France}
\author{Oladunjoye A.~Awoga}
\affiliation{Solid State Physics and NanoLund, Lund University, Box 118, S-221 00 Lund, Sweden}
\author{Nicholas Sedlmayr}
\affiliation{Institute of Physics, Maria Curie-Sk\l{}odowska University,
Plac Marii Sk\l{}odowskiej-Curie 1, PL-20031 Lublin, Poland}
\author{Annica M.~Black-Schaffer}
\affiliation{Department of Physics and Astronomy, Uppsala University, Box 516, S-751 20 Uppsala, Sweden}
\author{Catherine P\'epin}
\affiliation{Institut de Physique Th\'eorique, Universit\'e Paris Saclay, CEA
CNRS, Orme des Merisiers, 91190 Gif-sur-Yvette Cedex, France}
\author{Cristina Bena}
\affiliation{Institut de Physique Th\'eorique, Universit\'e Paris Saclay, CEA
CNRS, Orme des Merisiers, 91190 Gif-sur-Yvette Cedex, France}

\date{\today}

\begin{abstract}
We consider the most energetically favorable symmetry-allowed spin-singlet and spin-triplet superconducting pairing symmetries in monolayer and few-layer graphene, and for each calculate the energy spectrum in the presence of a scalar or magnetic impurity. We find that two doubly degenerate subgap states exist for scalar impurities for all types of pairing, except for the spin-singlet $s$-wave state. For magnetic impurities, two or four subgap states may form depending on the order parameter symmetry. We find that the spin polarization of these states allows one to distinguish between spin-singlet and triplet pairing, for example, only the spin-triplet states show opposite-energy subgap states with the same spin. We also calculate the quasi-particle interference patterns associated with the subgap states and find that they exhibit features that could distinguish between different types of pairing symmetries, especially a breaking of rotational symmetry for nodal states, stronger for the spin-singlet $d_{xy}$ and $d_{x^2-y^2}$ than for the spin-triplet $p_x$ and $p_y$ states.
\end{abstract}

\maketitle

\section{Introduction}
Ever since the discovery of superconductivity (SC) in graphene-based systems, such as twisted bilayer graphene \cite{cao2018unconventional,po2018origin,lu2019superconductors,balents2020superconductivity,andrei2020graphene,oh2021evidence,cao2021nematicity,christos2020superconductivity,chichinadze2020nematic,wu2020harmonic,yu2021nematicity,fischer2022unconventional} and rhombohedral trilayer graphene \cite{Zhou2021SuperconductivityIR, ghazaryan2021unconventional}, the identification of the pairing symmetries in these unconventional superconductors has been one of the main goals of the theoretical and experimental developments. However, the present state in the analysis of microscopic theories for the different graphene systems does not allow for a definite answer to this question \cite{lake2022pairing}. Multiple different mechanisms have been proposed based on both phonon-mediated pairing\cite{lian2019twisted,chou2021Acoustic} and electron-electron interactions \cite{kennes2018strong,jimeno2022superconductivity,pantaleon2022superconductivity} and they predict different pairing symmetries \cite{BlackSchaffer07, Nandkishore12, Kiesel12,vuvcivcevic2012d,BlackSchafferHonerkamp14,Awoga2017Domain,alidoust2019symmetry,alidoust2020josephson,thingstad2020phonon, chichinadze2020nematic, alsharari2022inducing, christos2020superconductivity,roy2019unconventional,roy2010unconventional,szabo2021extended,szabo2022competing,wolf2022topological,wolf2018unconventional}. 

In two recent works \cite{pangburn2022superconductivity,crepieux2022superconductivity} we have examined all expected spin-singlet and spin-triplet SC states  with lowest angular momentum ($l\le3$) in monolayer graphene, as well as in AB-stacked bilayer and ABA- and ABC-stacked trilayer graphene. Our goal was to analyze both the basic electronic properties and the topological properties of various SC graphene systems by computing their band structure and density of states (DOS), as well as the Chern number and the associated topologically protected edge states \cite{Hasan2010ColloquiumT,BlackSchafferHonerkamp14,Awoga2017Domain}. The analysis of these properties may help  to experimentally distinguish between various order parameters. For example, the DOS, measurable in scanning tunneling microscopy (STM), can in principle distinguish between nodal SCs (\mbox{$d_{xy}$-,} $d_{x^2-y^2}$-, \mbox{$p_x$-,} $p_y$-wave), which have a V-shaped DOS and gapped SCs ($s_{\rm on}$-, \mbox{$s_{\rm ext}$-,} $p+ip\,'$-, $d+ id$-, $f$-wave), which have a U-shaped DOS. However, in real experiments it can still be hard to distinguish between these two types of DOS if the resolution is not sufficient and disorder may additionally locally perturb the SC state \cite{Anderson1959TheoryOD}.

In this last work in the series, we propose another tool to distinguish between different SC order parameters in graphene by studying the effects of a single impurity on the local density of states (LDOS) and on the spin-polarized local density of state (SPLDOS). It is already well-known from the Anderson theorem~\cite{Anderson1959TheoryOD} that conventional $s$-wave SCs are not affected by scalar impurities, and do not allow the formation of subgap states, whereas scalar impurities usually induce subgap bound states/resonance states in unconventional fully gapped or nodal SCs~\cite{balatsky2006impurity}. Also, it is well-known that a magnetic impurity induces Yu-Shiba-Rusinov subgap states in a SC due to local time reversal symmetry breaking~\cite{Yu1965,Shiba1968,Rusinov1969}. In what follows we refer to all low-energy impurity states, irrespective of their origin, as simply subgap states. 

Although subgap states are often expected, their multiplicity and characteristics strongly depend on the underlying symmetries of the normal state and especially the SC order parameter~\cite{balatsky2006impurity,Awoga2018}.  
Thus, one may hope that studying the features induced by a single impurity would help differentiate between the different SC states. Motivated by these prospects, we perform an extensive study of the effects of both scalar and magnetic impurities in monolayer, AB bilayer, and trilayer ABA and ABC graphene for the same set of symmetries considered in Ref.~\onlinecite{pangburn2022superconductivity, crepieux2022superconductivity}.  Using the well-known T-matrix approach~\cite{balatsky2006impurity}, we compute both the spatially averaged LDOS as a function of energy, as well as the Fourier transform of the LDOS change induced by the impurity at a given energy (FT-LDOS, also known as the quasi-particle interference pattern, or QPI). The QPI and the spin-polarized QPI, measurable via STM and spin-polarized STM, respectively, provide a direct connection to angle-resolved photoemission spectroscopy experiments~\cite{SimonFTSTS2011,hoffman2002imaging,mcelroy2003relating,zhang2009experimental}, and thus contain information  about the band structure of the system. In fact, such measurements have already been used to study impurity scattering effects in graphene systems~\cite{bena2005quasiparticle,Bena2008effect,rutter2007scattering,BrihuegaQPI2008,Mallet2012role,Awoga2018,kaladzhyan2021quasiparticle1,joucken2021direct}.

We first focus on the energy dependence of the spatially averaged LDOS and SPLDOS. This allows us to study the formation of subgap states, thereby distinguishing between conventional and unconventional superconductors. For example, we confirm that in the presence of scalar impurities the spin-singlet $s$-wave states do not give rise to subgap states, according to the Anderson theorem \cite{Anderson1959TheoryOD,Awoga2018}. 
On the other hand, magnetic impurities are pair-breaking for both spin-singlet and spin-triplet states and thus generate subgap states for all types of pairing symmetries. Here we identify four different subgap states for the fully gapped $d+id\,'$, $p+ip\,'$ and $f$-wave symmetries, while the rest exhibiting only two subgap states.
For the SPLDOS generated by a magnetic impurity we note that pairs of subgap states of the same spin but opposite energies exist only for some of the spin-triplet order parameters but for no spin-singlet order parameters, which thus becomes a clear experimental signature allowing to identify the existence of a spin-triplet SC state. Another difference between the spin-singlet and spin-triplet SC states is that the SPLDOS features depend on the impurity spin orientation, while they are automatically independent of the direction of the impurity spin for all spin-singlet SC states. 
We subsequently study the QPI maps and show that the QPI patterns for nodal SC states break the six-fold symmetry of the normal state, while the QPI patterns for the gapped SC states preserve this symmetry. This establishes that the QPI can distinguish between gapless and nodal order parameters, such as \mbox{$p_x$-,} \mbox{$p_y$-,} \mbox{$d_{xy}$-,} $d_{x^2-y^2}$-wave, from fully gapped order parameters, such as $s$, \mbox{$s_{\rm ext}$-,} \mbox{$p+ip^{\,\prime}$-,} $d+id^{\,\prime}$- and $f$-wave,  in SC graphene systems. 
Finally, we note that most of the features of the subgap states are quite generic and unchanged when studying bilayer or trilayer graphene, except for extra subgap states arising in the ABC trilayer case and for a smaller splitting of the features due to the presence of the interlayer coupling.

The rest of this article is organized as follows. In Section \ref{sec:Model} we provide the details of the tight-binding model used and the T-matrix formalism. In Section \ref{sec:Monolayer} we focus on monolayer graphene and we compute the averaged LDOS and SPLDOS for both scalar and magnetic impurities, as well as their momentum dependence, or equivalently, the QPI. We extend the study to multilayer graphene in Section \ref{sec:Multilayer}, before summarizing our results in Section \ref{sec:conclusion}.

\section{Model and method\label{sec:Model}}
\subsection{Bulk Hamiltonian}
Without trying to justify the pairing mechanism for SC, we consider SC graphene described by a tight-binding Hamiltonian with a pairing term that can take all relevant spin-singlet and spin-triplet symmetries with the lowest angular momentum,
The non-interacting Hamiltonian is given by
\begin{eqnarray}
\begin{split}
\label{equation1}
H_0(\mathbf{k})=& \sum_{\mathbf{k},\alpha}	\mu\left(a_{\mathbf{k} \alpha}^\dagger a_{\mathbf{k} \alpha} + b_{\mathbf{k} \alpha}^\dagger b_{\mathbf{k} \alpha} \right) +
\\
&  h_{0}(\mathbf{k}) a_{\mathbf{k} \alpha}^\dagger b_{\mathbf{k} \alpha} + h_{0}^*(\mathbf{k}) b_{\mathbf{k} \alpha}^\dagger a_{\mathbf{k} \alpha},
	\label{h00}
 \\
h_{0}(\mathbf{k})&=-te^{-ik_{y}}\left[1+2e^{3ik_y/2}\cos\left(\dfrac{\sqrt{3}}{2}k_{x}\right)\right],
\end{split}
\end{eqnarray}
where $\mu$ and $h_{0}(\mathbf{k})$ are the chemical potential and the kinetic energy, respectively, with  $t$ denoting the hopping strength between nearest neighbor carbon atoms (NN). Here  $a_{\mathbf{k} \alpha}^\dagger$ $(b_{\mathbf{k} \alpha}^\dagger)$ is the creation operator for an electron with momentum $\mathbf{k}$ and spin $\alpha$, in the sublattice A (B).

We focus primarily on the intralayer NN SC pairing but our results are quite generic and are affected very little if we were to instead consider intralayer next-to-nearest-neighbor (NNN) order parameters, similarly to our earlier to works \cite{pangburn2022superconductivity,crepieux2022superconductivity,AwogaABC}. This is important to note since self-consistent calculations has shown that NNN range may be preferred over the NN for some multilayer graphene configurations \cite{chou2021Acoustic,AwogaABC}. The only type of pairing symmetry that cannot be captured by NN pairing is the $f$-wave state where we thus revert to NNN pairing. The SC Hamiltonian can be written as
\begin{equation}
H^0_\text{NN}= \sum \limits_{{\mathbf{k}}}h_\text{NN}^{0}({\mathbf{k}}) ( a^\dagger_{{\mathbf{k}}\up} b^\dagger_{-{\mathbf{k}}\down}  - a^\dagger_{{\mathbf{k}}\down}b^\dagger_{-{\mathbf{k}}\up}) + h.c.,
\end{equation}
and 
\begin{eqnarray}
H^x_\text{NN}=&\sum \limits_{{\mathbf{k}}} h_\text{NN}^x ({\mathbf{k}})(a^\dagger_{{\mathbf{k}}\up} b^\dagger_{-{\mathbf{k}}\up} -  a^\dagger_{{\mathbf{k}}\down} b^\dagger_{-{\mathbf{k}}\down} )+ h.c., \\
H^y_\text{NN}=& i\sum \limits_{{\mathbf{k}}}h_\text{NN}^y({\mathbf{k}})( a^\dagger_{{\mathbf{k}}\up} b^\dagger_{-{\mathbf{k}}\up}+ a^\dagger_{{\mathbf{k}},\down} b^\dagger_{-{\mathbf{k}}\down}) + h.c., \\
H^z_\text{NN}=& \sum \limits_{{\mathbf{k}}}h_\text{NN}^z({\mathbf{k}})(a^\dagger_{{\mathbf{k}}\up}b^\dagger_{-{\mathbf{k}}\down}+ a^\dagger_{{\mathbf{k}}\down}b^\dagger_{-{\mathbf{k}}\up})+ h.c. ,
\label{HSCtriplet}
\end{eqnarray}
for the spin-singlet channel and spin-triplet channels, respectively \cite{pangburn2022superconductivity,crepieux2022superconductivity,AwogaABC}.
Here $h_\text{NN}^{\eta}({\mathbf{k}})$ are the overall form factors whose expressions depend on both the spin channel chosen and the angular momentum symmetry of the order parameter. Their expressions for the different pairing symmetries are given in Table~\ref{table1}. For the NNN range the above formulas need to be modified such that the pairing terms couple two electrons within the same sublattice. The NNN form factor for the $f_x = f_{x(x^2-y^2)}$ order parameter, which is the only one considered here, is also given in Table~\ref{table1} (we exclude the $f_{y(y^2-3x^2)}$-wave state has it has multiple nodes and is as such highly unfavorable).

\begin{table}
\renewcommand\arraystretch{2}
\resizebox{8.5cm}{!}{
\begin{tabular}{|c|l|l|}
  \hline
  $\quad\eta\quad$ & Symmetry & Form factor $h_\text{NN}^{\eta}(\mathbf{k})$ \\
  \hline
 \;$0$ & \;$s_\text{ext}$ & $ h_\text{NN}^{0,s_\text{ext}}({\mathbf{k}})= \frac{\Delta_0}{\sqrt{3}} \tilde{h}_0({\mathbf{k}})$  \\
 \;$0$ & \; $d_{x^2-y^2}$ & $ h_\text{NN}^{0,d_{x^2-y^2}}({\mathbf{k}})= \frac{2\Delta_0}{\sqrt{6}}e^{-ik_y}\left[1-e^{\frac{3i}{2}k_y}\cos(\frac{\sqrt{3}}{2}k_x)\right]$ \\
\;$0$ & \; $d_{xy}$ & $ h_\text{NN}^{0,d_{xy}}({\mathbf{k}})=  \Delta_{0}\sqrt{2}  i \ e^{\frac{i}{2}k_y}\sin(\frac{\sqrt{3}}{2}k_x)$ \\ 
  \hline
 \; $x$ & \;$p_y$ & $h_\text{NN}^{\eta,p_y}({\mathbf{k}})= \frac{2 \Delta_0}{\sqrt{6}}e^{-ik_y}\left[1-e^{\frac{3i}{2}k_y}\cos(\frac{\sqrt{3}}{2}k_x)\right]$  \\
 \; $x$ & \; $p_x$&  $h_\text{NN}^{\eta,p_x}({\mathbf{k}})= i\sqrt{2}\Delta_0 e^{\frac{i}{2}k_y}\sin(\frac{\sqrt{3}}{2}k_x)$ \\
 \; $x$ & \; $f_{x(x^2-y^2)}$ & $ h_\text{NNN}^{\eta,f_x}({\mathbf{k}})=\frac{2i\Delta_0}{\sqrt{6}}\Big[\sin(\sqrt{3}k_x)$ \\
 && \hspace*{1.5cm}$-2\sin(\frac{\sqrt{3}}{2}k_x) \cos(\frac{3}{2}k_y)\Big]$ \\
  \hline
\end{tabular}}
\caption{Expressions for the SC form factors for different spin-singlet and spin-triplet symmetries. The overall amplitude for the SC order parameter is set to $\Delta_0$, the distance between two NN carbon atoms is set to 1, and $\tilde{h}_0({\mathbf{k}})=h_0({\mathbf{k}})/t$.}
\label{table1}
\end{table}

For SC multilayer graphene, the Hamiltonian is given by
\begin{equation}
H_{\mathbf{k}}=\sum_{\ell=1}^L \left(H_0^{(\ell)}+H_\text{NN}^{(\ell)}\right)+H_\text{inter-layer}~,
\end{equation}
where $L$ is the number of layers, $H_0^{(\ell)}$ and $H_\text{NN}^{(\ell)}$ are the non-interacting and the SC Hamiltonians, respectively, associated to each layer $\ell$ and given by Eqs.~(\ref{h00})--(\ref{HSCtriplet}), and $H_\text{inter-layer}$ is the coupling Hamiltonian between adjacent layers given in Ref.~\onlinecite{pangburn2022superconductivity}. The interlayer Hamiltonian depends on three additional parameters, the phase difference $\phi$ between the SC state in two adjacent layers, the interlayer couplings $\gamma_1$ and  $\gamma_3$\cite{malard2007probing}, where $\gamma_1$ is the simple inter-layer coupling corresponding to hopping between atoms on top of each other, while the smaller  $\gamma_3$ corresponds to hopping between an atom $A$ in one layer and the neighboring $B$ atoms in the adjacent layer.  We have checked that the addition of this trigonal warping $\gamma_3$ with a realistic value $\gamma_3 \le \gamma_1$ does not change our results. We thus set $\gamma_1=0.2 t$ and $\gamma_3=0$ in the rest of the work for simplicity.

Collating the operators in each layer $\ell$  into a vector, the Hamiltonian can be expressed as
\begin{equation}
H_{\mathbf{k}}=\frac{1}{2} \Psi_{\mathbf{k}}^{\dagger}  \hat H_{\mathrm{BdG}} \Psi_{\mathbf{k}},
\end{equation} 
using the basis
\begin{equation}\label{eq:Basis}
\Psi_{\mathbf{k} \ell }=\left(a_{\mathbf{k} \ell \uparrow}, b_{\mathbf{k} \ell \uparrow}, a_{\mathbf{k} \ell \downarrow}, b_{\mathbf{k} \ell \downarrow}, a_{-\mathbf{k} \ell \uparrow}^{\dagger}, b_{-\mathbf{k} \ell \uparrow}^{\dagger}, a_{-\mathbf{k} \ell \downarrow}^{\dagger}, b_{-\mathbf{k} \ell \downarrow}^{\dagger}\right)^T,
\end{equation}
where $\Psi_{\mathbf{k}}$ thus combines all individual-layer bases $\Psi_{\mathbf{k} \ell }$, and $\hat H_{\mathrm{BdG}}$ is the $8 L \times 8 L$ Bogoliubov-de-Gennes (BdG) Hamiltonian matrix. The factor 8 corresponds to a product of 2 spins, 2 sublattices, and the particle-hole doubling of the degrees of freedom in the BdG formalism. 
Finally, The retarded Green's function for this system is given by
\begin{equation}\label{eq:G0}
G^{\,r}(E,\mathbf{k})=\left[ E+i\delta- \hat H_{\mathrm{BdG}}(\mathbf{k})\right]^{-1},
\end{equation}
with $\delta$ being the quasiparticle-lifetime. We set $\delta=0.03$ in the rest of the work.

\subsection{Impurity scattering}

In this work we are interested in the consequences of introducing a point-like (scalar or magnetic) impurity. Using the basis in Eq.~\eqref{eq:Basis}, the Hamiltonian matrix for such point-like impurity can be written as
\begin{equation}
	\hat {\mathds{V}} = \tau^z \otimes \hat V, \qquad \hat V =\hat u \otimes \hat v.
\end{equation}
Here
\begin{equation}
	\hat v =
	U\sigma^0+J\sigma^\nu,
\end{equation}
where $\tau^\nu (\sigma^\nu) $ are the $\nu$-Pauli matrices in the particle-hole (spin) space, and $\sigma^0$ is the $2\times 2$ identity matrix, while $ \hat u$ is a $2L \times 2L$ matrix for which all the elements are equal to $0$, except for one diagonal element, whose matrix position $i_{\rm imp}$ corresponds to the layer/sublattice of the impurity, and which we take to be equal to $1$.  The parameters $U$ and  $J$ are, respectively, the strength of the scalar and magnetic impurity.

To compute the corresponding variation of the unpolarized (LDOS) and spin-polarized (SPLDOS) local density of states, we use the T-matrix approach\cite{balatsky2006impurity}. The T-matrix can be written as
\begin{equation}
\text{T}(E)=\left[\mathds{1}_{8 L}-\hat {\mathds{V}}  \int\dfrac{d^2\mathbf{k}}{(2\pi)^2} G^{\,r}(E,\mathbf{k})\right]^{-1}\hat {\mathds{V}},\label{eq:Tmatrix}
\end{equation}
where $\mathds{1}_{8 L}$ is an ${8 L} \times {8 L}$ identity matrix.
The physical observables (here LDOS and SPLDOS), that can be measured near an impurity, can be expressed directly in terms of this T-matrix, if we assumes the dilute-limit approximation, such that the impurities are well separated from each other. The Fourier transform of the change in LDOS induced by the impurity, $\delta \rho (\mathbf{q},E)$, and the same quantity for the SPLDOS,  $\delta S_{\nu}(\mathbf{q},E)$, can then be written as
\begin{eqnarray}\label{eq:QPI}
&&\delta \rho(\mathbf{q},E)=-\dfrac{1}{2\pi i}\int \dfrac{d^2 \mathbf{k}}{(2\pi)^2}\nonumber\\
&&\times\sum\limits_{b}\left[\,\tilde{g}_{b,\uparrow \uparrow}(E,\mathbf{q},\mathbf{k})+\tilde{g}_{b,\downarrow \downarrow}(E,\mathbf{q},\mathbf{k}) \,\right],\\
&&\delta  S_{x}(\mathbf{q},E) =-\dfrac{1}{2\pi i}\int \dfrac{d^2 \mathbf{k}}{(2\pi)^2}\nonumber\\
&&\times\sum\limits_{b}\left[\,\tilde{g}_{b,\uparrow \downarrow}(E,\mathbf{q},\mathbf{k})+\tilde{g}_{b,\downarrow \uparrow}(E,\mathbf{q},\mathbf{k}) \,\right],\\ %&\\[5pt]
&&\delta  S_{y}(\mathbf{q},E)=-\dfrac{1}{2\pi}\int \dfrac{d^2 \mathbf{k}}{(2\pi)^2}\nonumber\\
&&\times\sum\limits_{b}\left[\,g_{b,\uparrow \downarrow}(E,\mathbf{q},\mathbf{k})-g_{b,\downarrow \uparrow}(E,\mathbf{q},\mathbf{k})\, \right],\\ %&\\[5pt]
&&\delta  S_{z}(\mathbf{q},E)=-\dfrac{1}{2\pi i}\int \dfrac{d^2 \mathbf{k}}{(2\pi)^2}\nonumber\\
&&\times\sum\limits_{b}\left[\,\tilde{g}_{b,\uparrow \uparrow}(E,\mathbf{q},\mathbf{k})-\tilde{g}_{b,\downarrow \downarrow}(E,\mathbf{q},\mathbf{k})\, \right],
\label{eq:QPIend}
\end{eqnarray}
where the index $b$ runs over all electron bands (the hole bands are not taking into account since experimentally only the available electron density of states is measured) and 
\begin{equation}
\begin{split}
\label{eq_LDOS}
g/\tilde{g}(E,\mathbf{q},\mathbf{k})=&G^{\,r}(E,\mathbf{q})T(E)G^{\,r}(E,\mathbf{q+k})  \\
 & \pm (G^{\,r}(E,\mathbf{k+q}))^{\ast}T^{\ast}(E)(G^{\,r}(E,\mathbf{q}))^{\ast}.
\end{split}
\end{equation}
At $\mathbf{q}=0$, the quantities $\delta \rho(\mathbf{q}=0,E) \rightarrow \delta \rho(E)$ and $\delta  S_{\nu }(\mathbf{q}=0,E)\rightarrow \delta  S_{\nu }(E)$ correspond to the spatially averaged disorder-induced LDOS and SPLDOS, respectively. In the two following Sections, we plot and analyze $\delta \rho(E)$ and $\delta  S_{\nu }(E)$ as a function of energy and impurity strength to establish the formation of subgap states.
Furthermore, at constant energy, the QPI patterns described by Eq.~\eqref{eq:QPI}-\eqref{eq:QPIend} provide a map in reciprocal space of the possible scattering processes. Experimentally, the QPI patterns are obtained by performing a fast Fourier transform of the STM measurements of the LDOS in real space~\cite{BrihuegaQPI2008,Awoga2018}.

\section{Monolayer graphene\label{sec:Monolayer}}

\subsection{Unpolarized and spin-polarized local density of states}

We first consider the spatially averaged LDOS, i.e~$\delta \rho(E)$ and  $\delta  S_{\nu }(E)$, in the presence of both a scalar and a magnetic impurity. If and when subgap states form, these quantities will display clear peaks inside the SC gap. The position of the peaks may depend on various parameters, such as the impurity strength, the amplitude of the SC order parameter and its symmetry, or the chemical potential. 

\begin{figure*}[!htb]
\begin{center}
\resizebox{\linewidth}{!}{
    \begin{tabular}{ c  c  c  c  c }
      \thead{ \includegraphics[width=3.7cm]{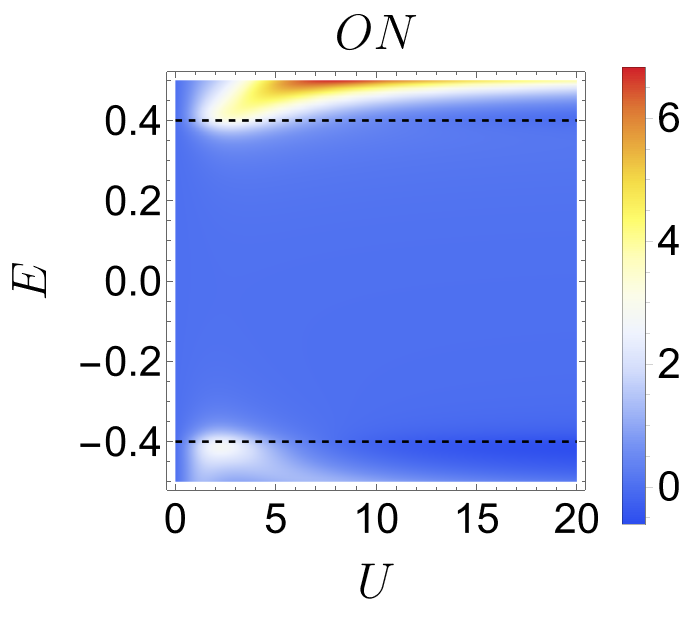}}
      &\thead{\includegraphics[width=4cm]{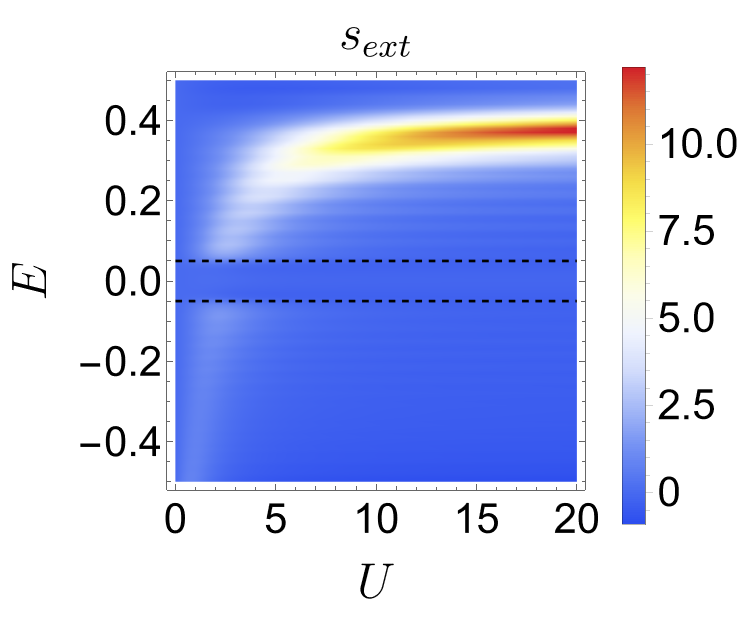}}
      &\thead{\includegraphics[width=4cm]{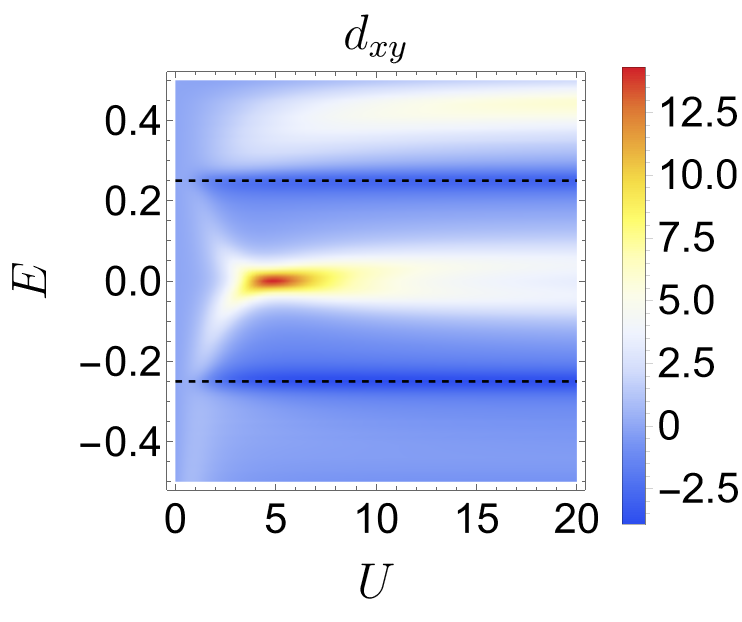}}
      & \thead{\includegraphics[width=4cm]{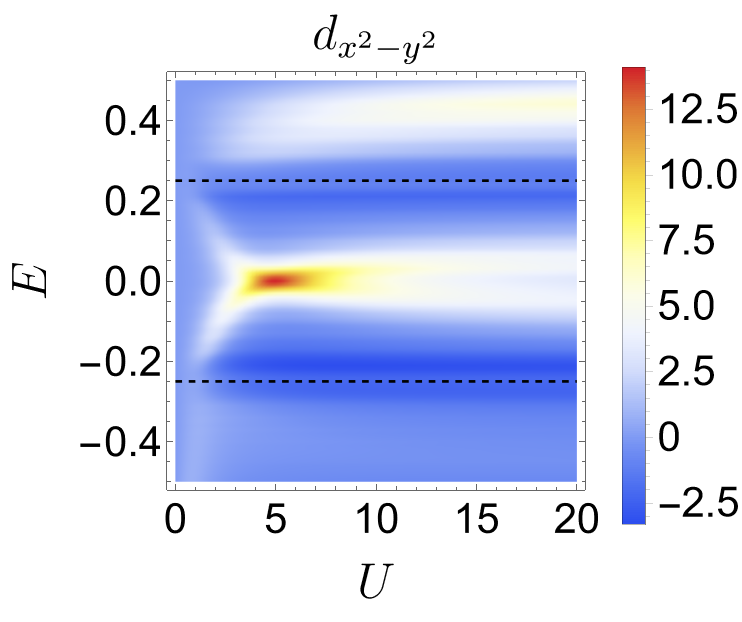}}
    &\thead{\includegraphics[width=3.9cm]{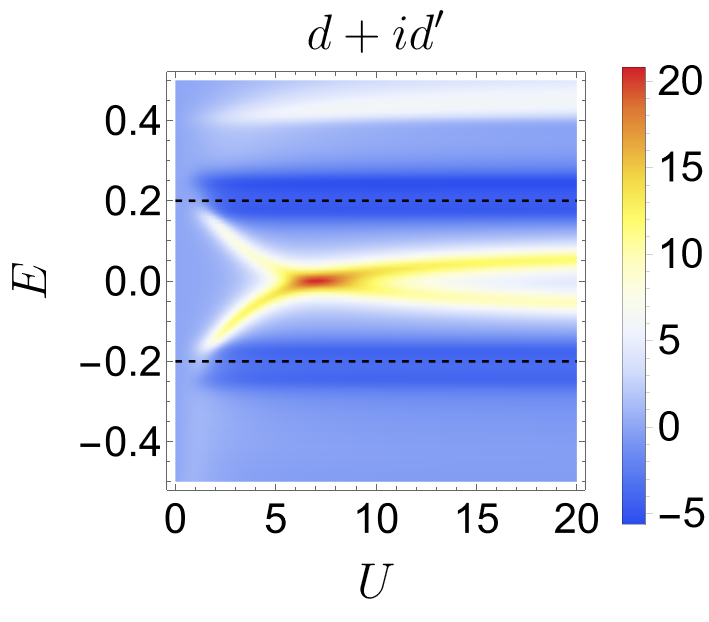}}     
    \vspace{-0.4cm}
    \\
      \thead{\includegraphics[width=4cm]{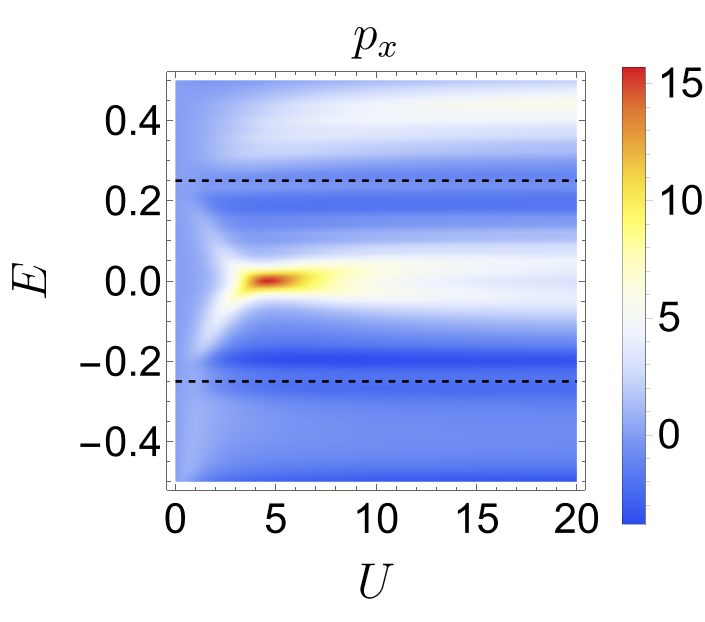}}
      & \thead{ \includegraphics[width=4.2cm]{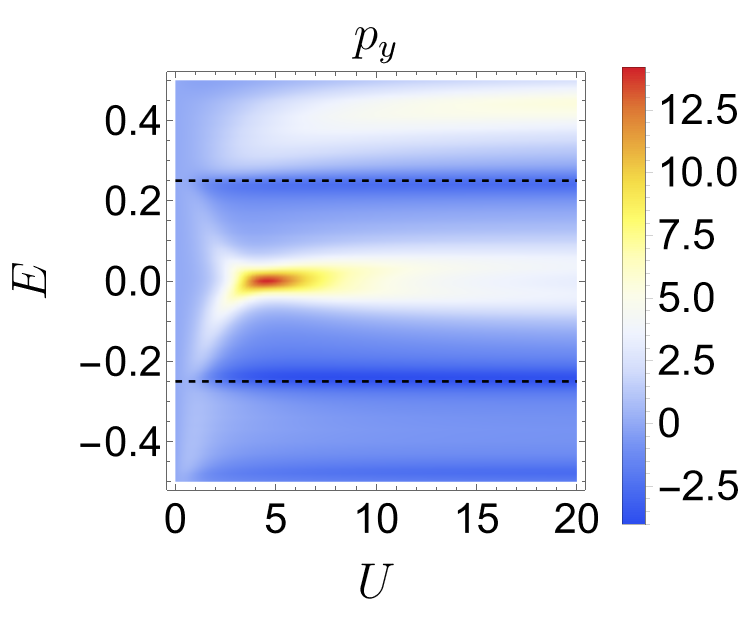} } 
      &\thead{\includegraphics[width=3.9cm]{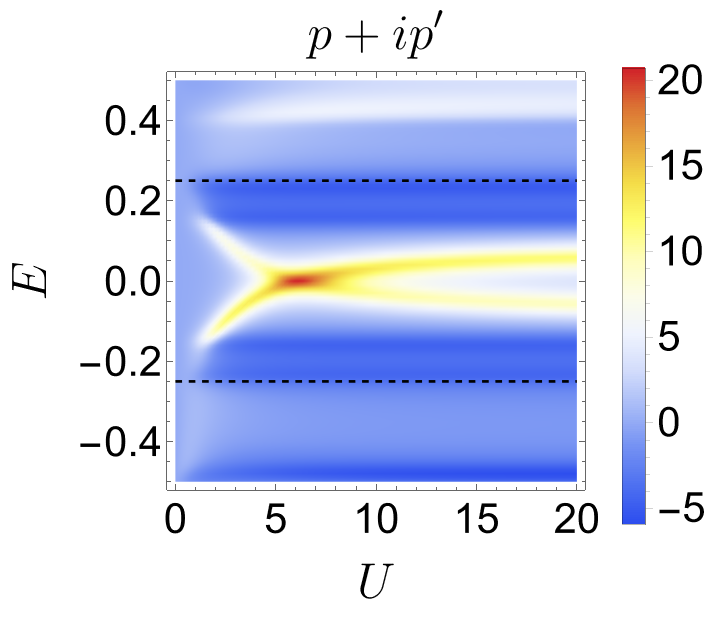} } 
      & \thead{\includegraphics[width=4cm]{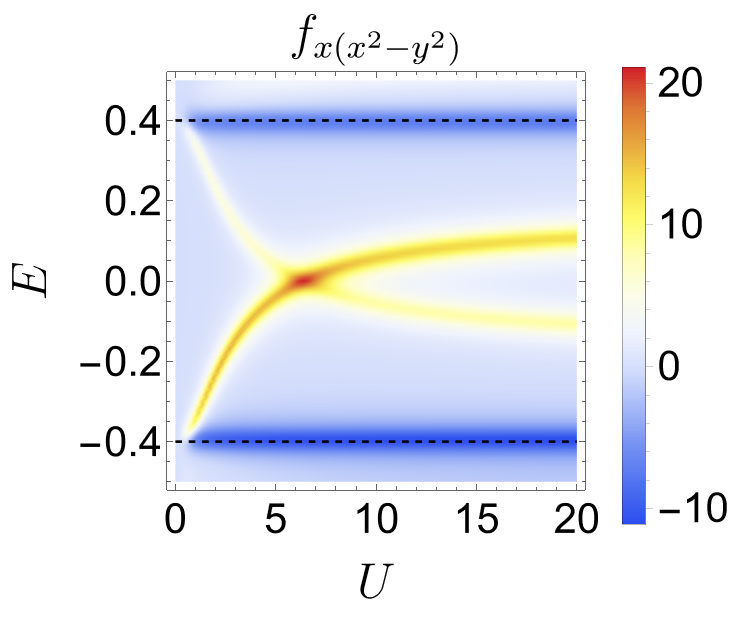} }
    \end{tabular}
    }
\vspace{-0.7cm}
\end{center}
\caption{$\delta \rho (E)$ as a function of energy $E$ and impurity strength $U$ for a scalar impurity. We take $\mu = 0.4t$ and $\Delta_0 = 0.4t$. The dotted lines indicate the SC gap edge, which as noted in Ref.~\onlinecite{pangburn2022superconductivity}, does not always lie at an energy equal to $\Delta_0$ but may depend on various parameters, including the symmetry of the SC order parameter.}
\label{mono_scal_rho}
 \end{figure*}

In Fig.~\ref{mono_scal_rho} we plot $\delta \rho (E)$ for all the SC symmetries as a function of energy and impurity strength in the presence of a scalar impurity. 
We first note that for the on-site (ON) and $s_{ext}$ symmetries, there is no impurity subgap peak. This is consistent with previous observations in the literature: conventional $s$-wave superconductors are unaffected by the presence of non-magnetic or scalar impurities \cite{Anderson1959TheoryOD,balatsky2006impurity,Awoga2018} and thus do not show any subgap states since these impurities do not break time reversal symmetry \cite{skvortsov_subgap_2013}. This reasoning can also be applied to extended $s$-wave superconductors: as long as the chemical potential is chosen such that the SC order parameter is almost constant along the Fermi surface, the phenomenology is approximately the same \cite{salkola1996theory}.

In contrast, the spin-singlet nodal, $d_{xy}$- and $d_{x^2-y^2}$-wave, as well as the fully gapped, chiral $d_{xy}+id_{x^2-y^2}$-wave ($d+id') $states, show spin degenerate subgap states in the presence of a scalar impurity. For these states the Anderson theorem does not forbid the presence of subgap states, even for a non-magnetic impurity. The physical interpretation is that scattering by an impurity disturbs the phase distribution for some particular directions of the momenta in all these nontrivial SC states \cite{balatsky2006impurity}. This has also been noted in $d$-wave superconductors on the square lattice, modeling the cuprate superconductors \cite{balatsky2006impurity} and is also in agreement with former studies on the chiral $d+id\,'$-wave SC in graphene \cite{lothman2014defects,Awoga2018}.
Similarly, we find subgap states for all spin-triplet states, both the nodal $p_x$- and $p_y$-wave states, and the fully gapped, chiral $p_x+ip_y$-wave (p+ip') SC state, as well as the fully gapped $f_x$-wave state.
For all these subgap states we find that their energies evolve with the impurity strength such that the states cross zero energy at a given, but different, impurity strength. A similar observation has also been noted in Ref.~\onlinecite{Kaladzhyan2016CharacterizingUS}.

\begin{figure*}[!htb]
\begin{center}
%\resizebox{1.1\columnwidth}{!}{%
\resizebox{\linewidth}{!}{%
    \begin{tabular}{ c  c  c  c  c }
    \thead{ \includegraphics[width=4.2cm]{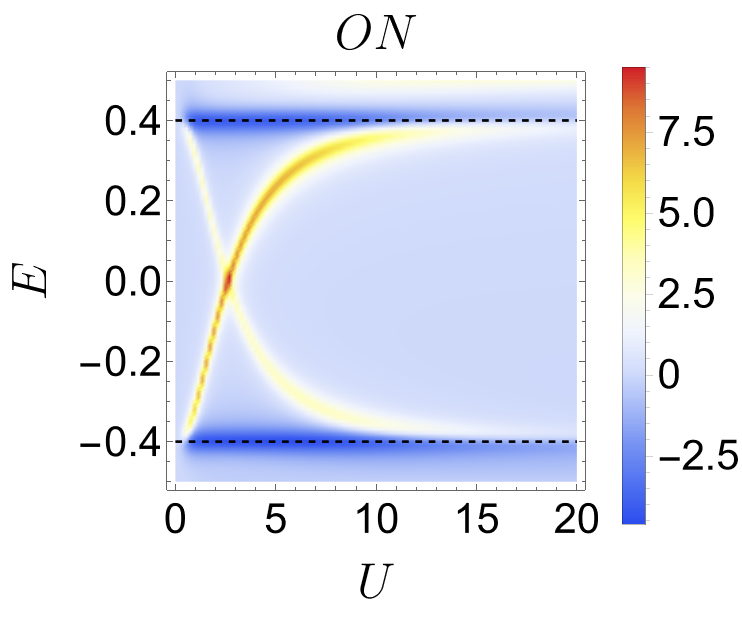}}
      & \thead{\includegraphics[width=4cm]{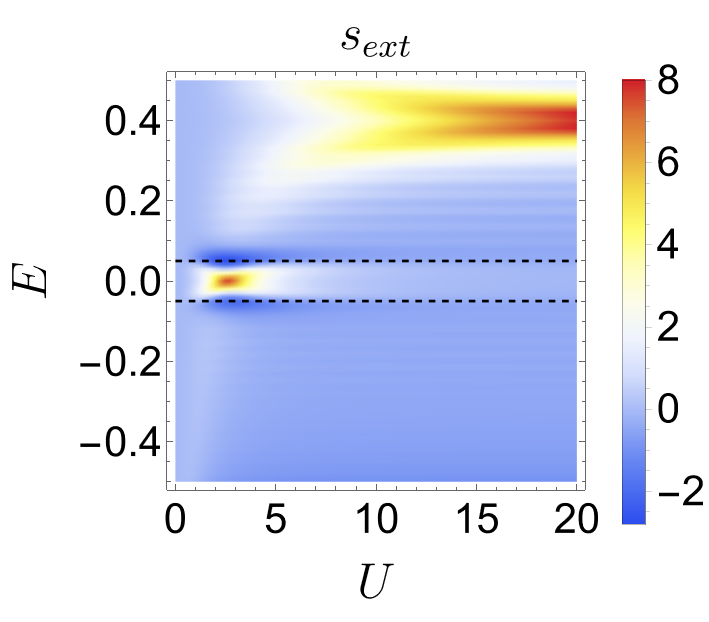}}
  & \thead{\includegraphics[width=4cm]{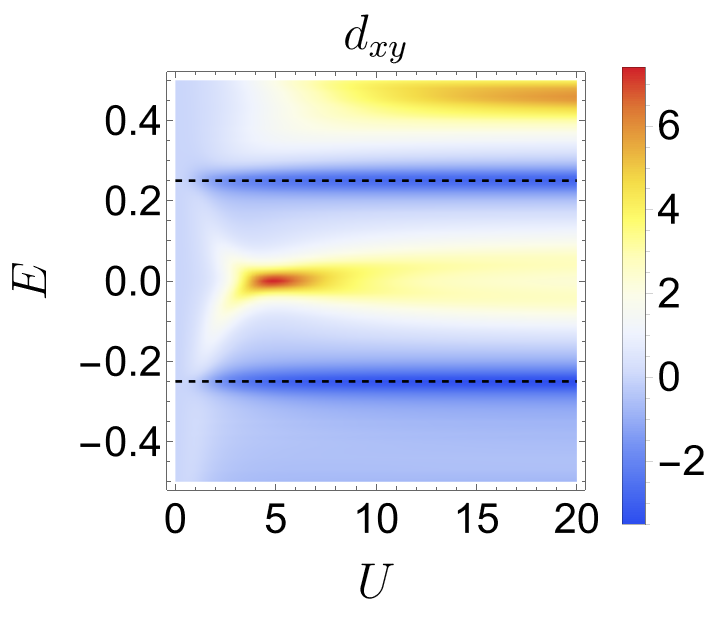}} 
   &\thead{\includegraphics[width=4cm]{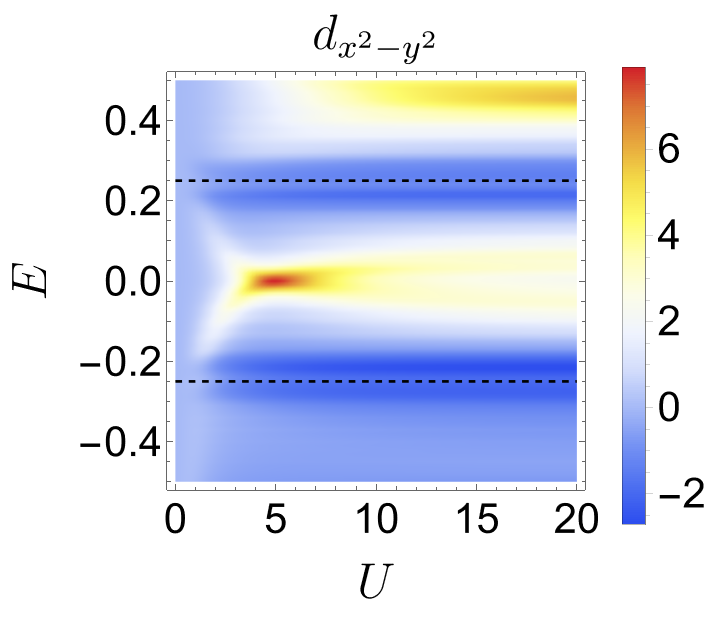}}
     &\thead{\includegraphics[width=4cm]{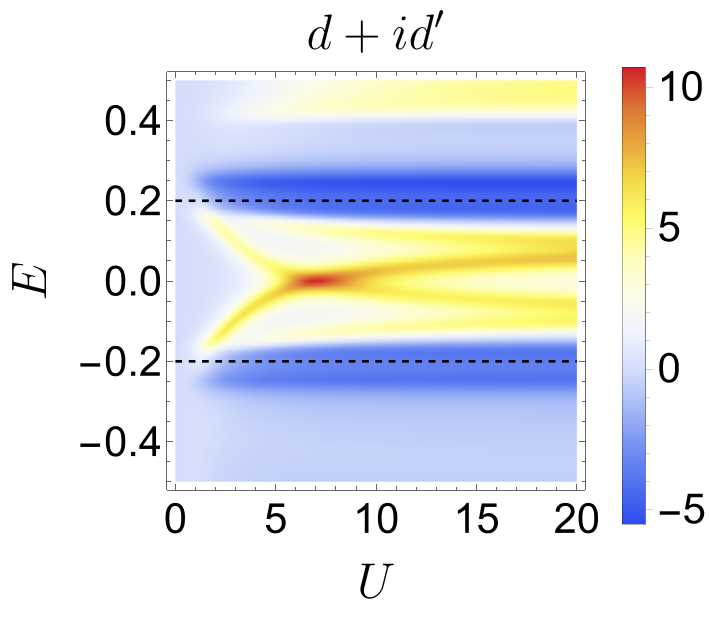}}
    \vspace{-0.4cm}
   \\
    \thead{\includegraphics[width=4cm]{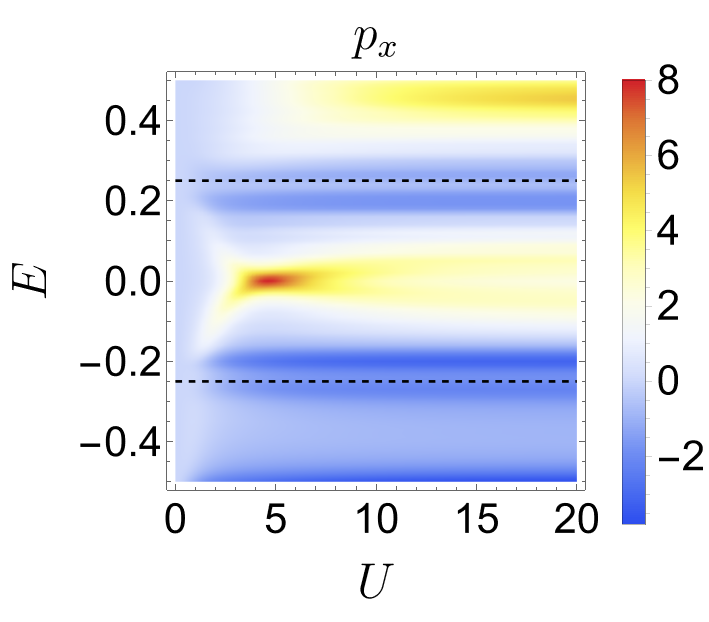}}
      & \thead{ \includegraphics[width=4cm]{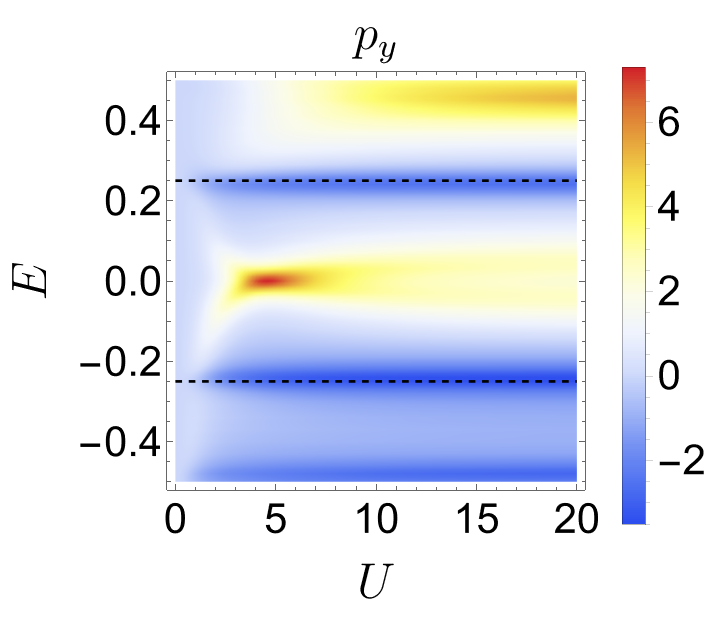} }
      &\thead{\includegraphics[width=4cm]{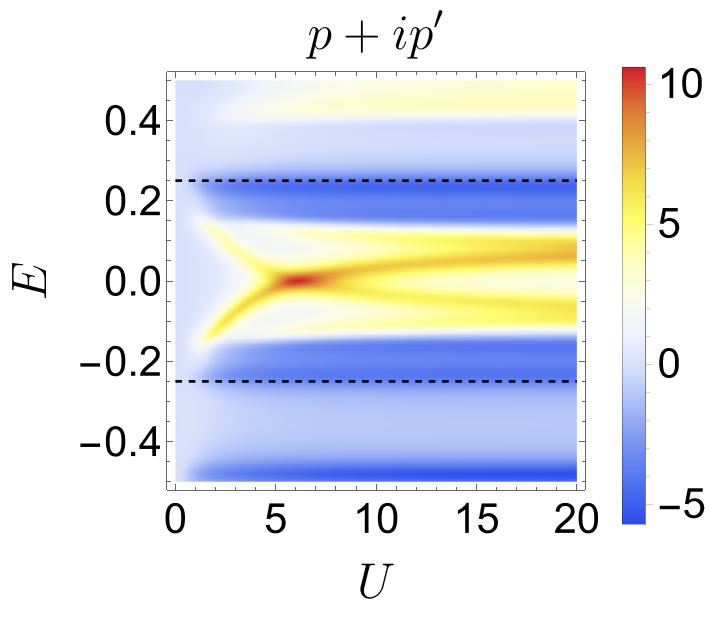} }
       &\thead{\includegraphics[width=4.2cm]{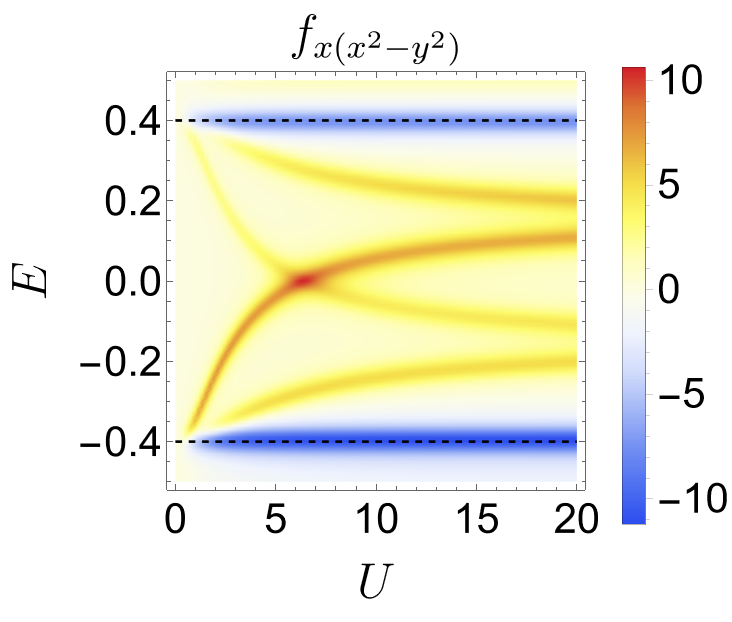} }
    \end{tabular}
    }%
    \vspace{-0.7cm}
  \end{center}
   \caption{ $\delta \rho (E)$ as a function of energy $E$ and impurity strength $J_z$ for a $z$-magnetic impurity. We take $\mu = 0.4t$ and $\Delta_0 = 0.4t$. The dotted lines indicate the gap edge.}
  \label{mono_magz_rho}
 \end{figure*}
 \begin{figure*}[!htb]
\begin{center}
%\resizebox{1.1\columnwidth}{!}{%
\resizebox{\linewidth}{!}{%
    \begin{tabular}{ c  c  c  c  c }
      \thead{ \includegraphics[width=4cm]{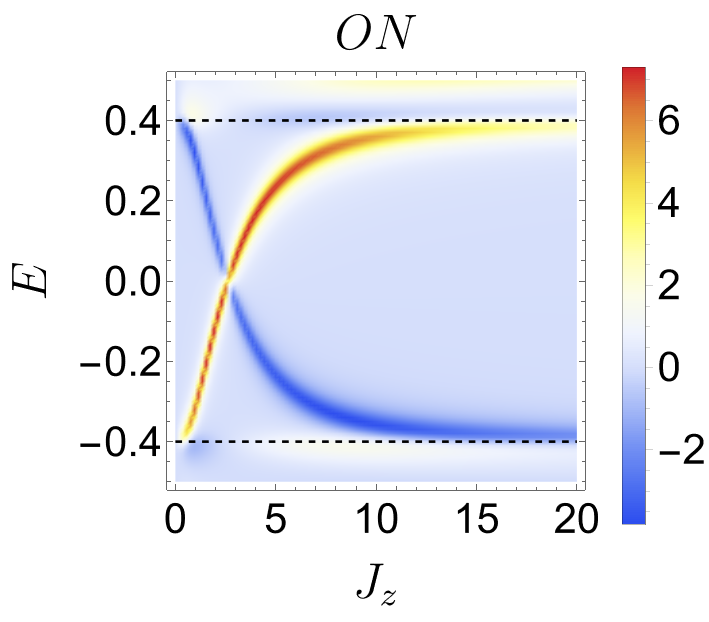}}
      & \thead{\includegraphics[width=4cm]{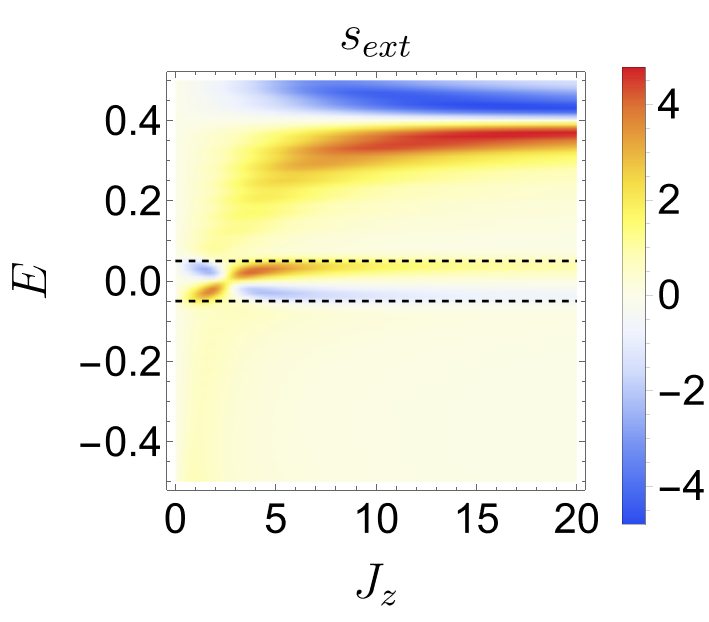}}
   & \thead{\includegraphics[width=4cm]{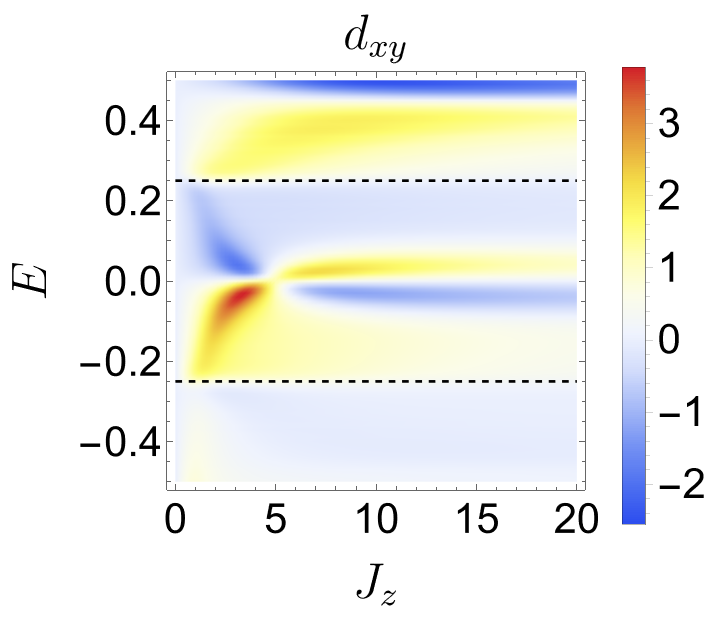}} 
   & \thead{\includegraphics[width=4cm]{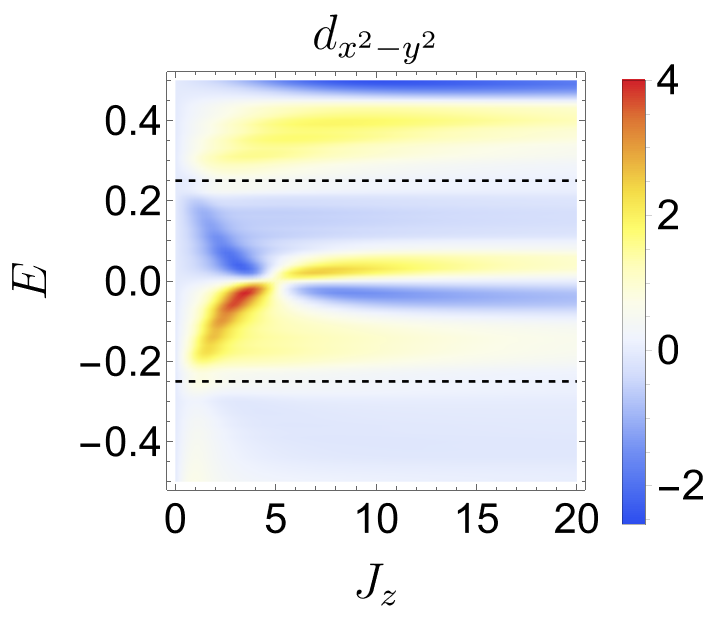}}
    &  \thead{\includegraphics[width=4.2cm]{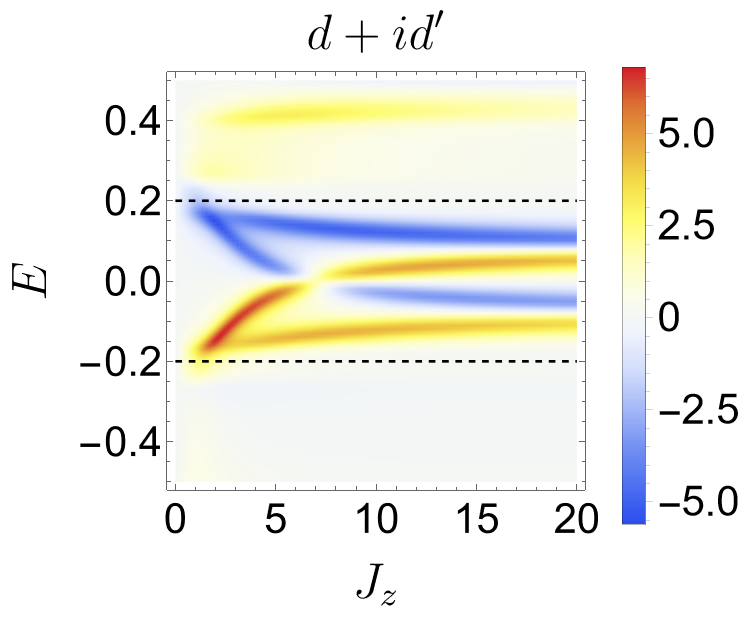}} 
       \vspace{-0.4cm}
   \\
     \thead{\includegraphics[width=4cm]{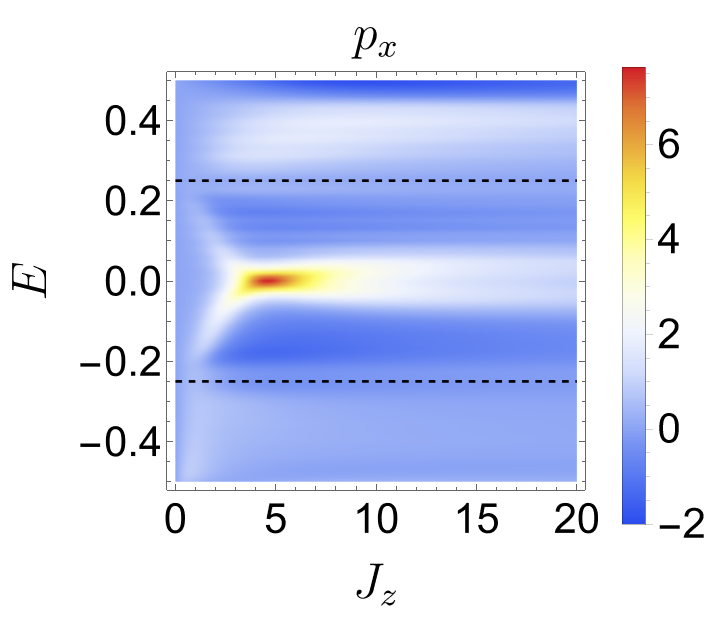}}
      & \thead{ \includegraphics[width=4cm]{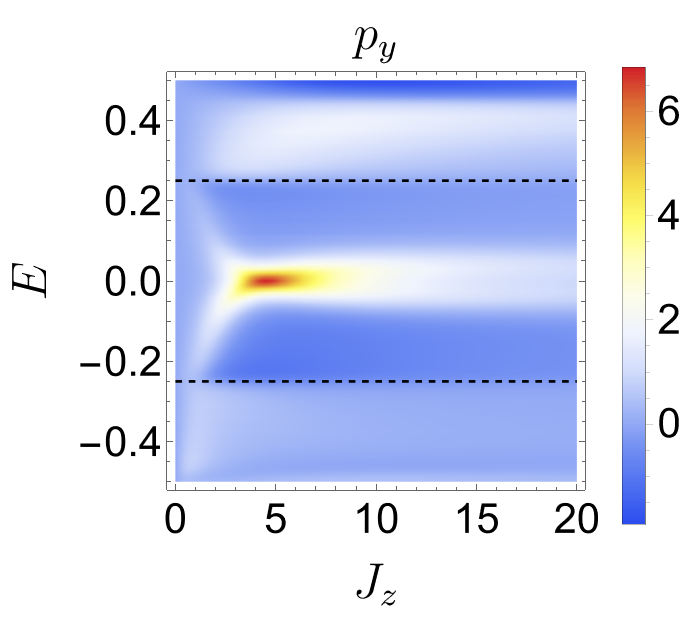} } 
       & \thead{\includegraphics[width=4.3cm]{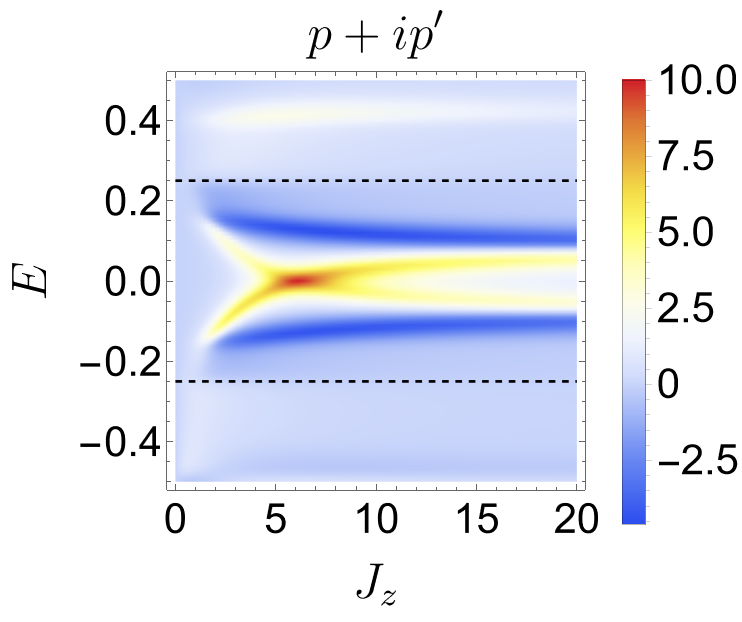} }
      & \thead{\includegraphics[width=4.2cm]{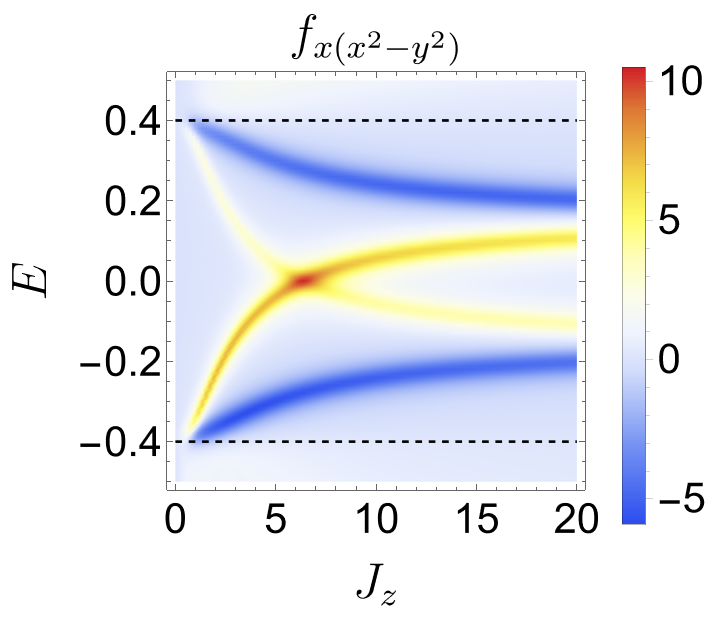} }
    \end{tabular}
    }%
    \vspace{-0.7cm}
  \end{center}
  \caption{ $\delta S_z(E)$ as a function of energy $E$ and impurity strength $J_z$ for a $z$-magnetic impurity. We take $\mu = 0.4t$ and $\Delta_0 = 0.4t$. The dotted lines indicate the gap edge.}
  \label{mono_magz_rho_and_Sz}
 \end{figure*}
 
We next study the effect of a magnetic impurity for the formation of subgap states. First we plot in Fig.~\ref{mono_magz_rho} $\delta  \rho(E)$ as a function for energy and impurity strength for all the order-parameter symmetries considered. We first note that for a magnetic impurity we find subgap states for all types of pairing symmetries, including the $s$-wave states. We also note that for the fully gapped $d+id\,'$-, $p+ip\,'$-, and $f$-wave states, the spin degeneracy has been lifted, and we have now four distinct subgap states rather than two pairs of degenerate ones. On the other hand, the nodal SC states $d_{xy}$-, $d_{x^2-y^2}$-, $p_x$-, and $p_y$-wave, show the same number of subgap states, i.e~two distinct states, for both magnetic and scalar impurities, however a pair of extra impurity states often arises outside the gap. As a consequence, the number of subgap states can be used to simple tool to experimentally distinguish between various order parameters, i.e.~if we can identify four distinct subgap states, then we can be sure to have either a $d+id\,'$-, $p+ip\,'$-, or $f$-wave pairing. We note however, that the reverse may not always work since the four states may be too close together to distinguish experimentally. 

\begin{figure}[!htb]
\begin{center}
\resizebox{0.8\linewidth}{!}{%
    \begin{tabular}{ c  c  c }
      \thead{\includegraphics[width=4cm]{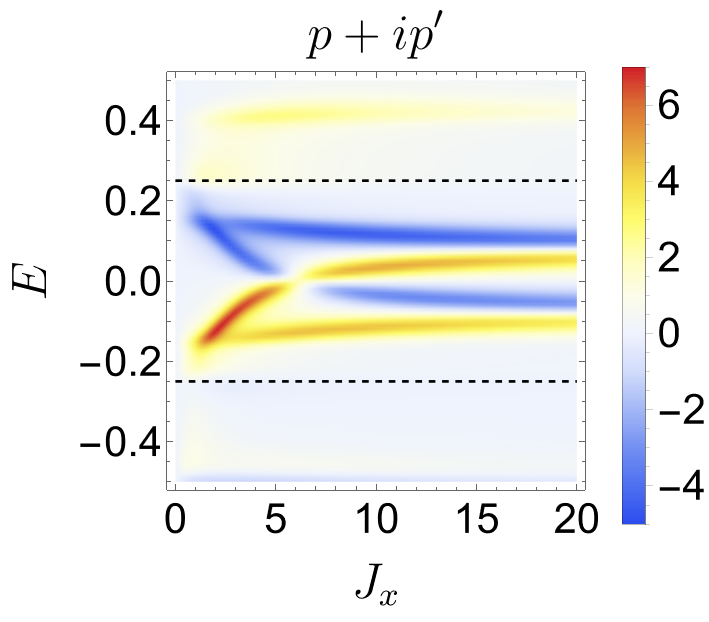}} 
   & \thead{
   \includegraphics[width=4cm]{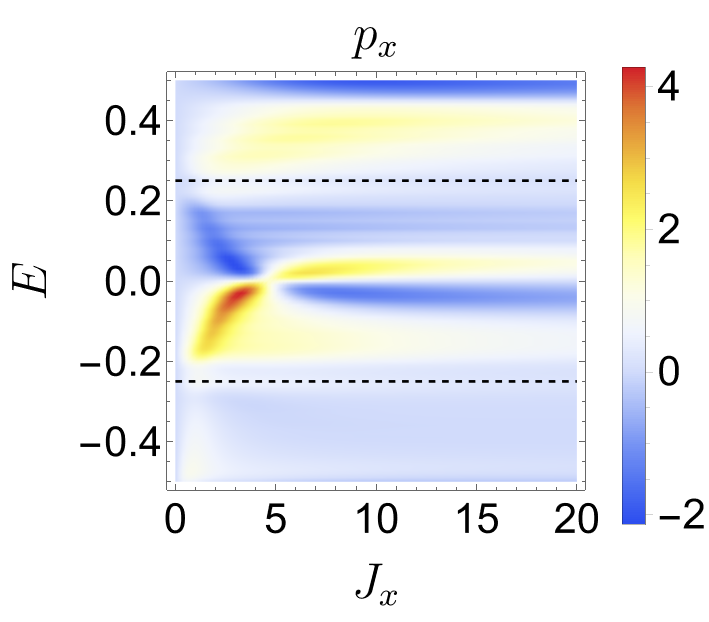}} 
    \vspace{-0.4cm}
   \\
      \thead{\includegraphics[width=4cm]{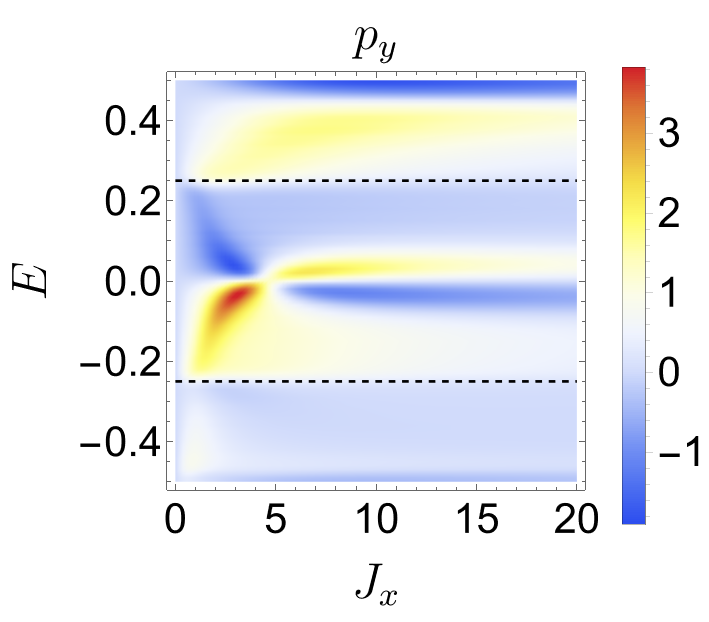} } 
      &  \thead{\includegraphics[width=4cm]{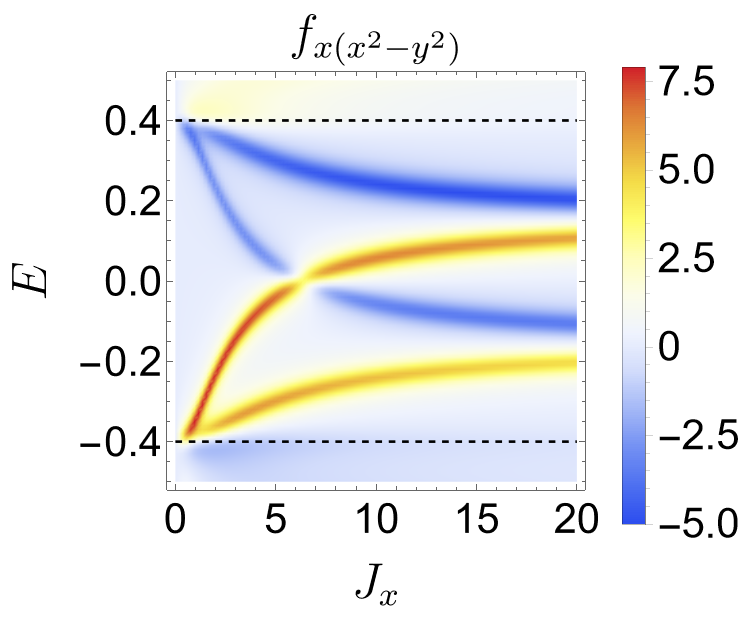}}  
      \\
    \end{tabular}
    }%
        \vspace{-0.7cm}
  \end{center}
  \caption{ $\delta S_x (E)$  as a function of energy $E$ and impurity strength $J_x$ for a $x$-magnetic impurity. We take $\mu = 0.4 t$ and $\Delta_0 = 0.4t$. The dotted lines indicate the gap edge.}
  \label{mono_Shiba_Sx_magx}
 \end{figure}

In order to get a better understanding of what happens for magnetic impurities, we also look at the spin-polarization of the induced subgap states. Here we find that each impurity gives rise to a non-zero spin polarization only in the spin channel parallel to its spin direction, thus for an $\alpha$-magnetic impurity we plot only the $\alpha$-magnetic component of the SPLDOS, with $\alpha=x,y,z$. 
We thus first plot in Fig.~\ref{mono_magz_rho_and_Sz} the SPLDOS $\delta S_z(E)$ as a function of magnetic impurity strength and energy for all types of pairing for a $z$-magnetic impurity. 
Here we find that the SPLDOS shows even more clearly the difference between the two and four subgap scenarios, as it clearly differentiates between the different states. We can next ask if the interplay between the direction of the impurity spin and the choice of the triplet channel, which we above fixed to $x$, influences the results. We have checked that for this particular choice a $y$-magnetic impurity yields exactly the same behavior as the $z$-magnetic impurity. However, for an $x$-magnetic impurity, the LDOS is unchanged but the $x$-SPLDOS differs in the spin-triplet channel. In Fig.~\ref{mono_Shiba_Sx_magx} we illustrate this by plotting the $x$-SPLDOS for all spin-triplet states.
 
We further note in Fig.~\ref{mono_magz_rho_and_Sz} that for the spin-singlet SC states, opposite-energy states have opposite spin, while for the spin-triplet states the opposite energy states have the same spin. This however seems to be a feature dependent on the direction of the impurity spin: for the different impurity direction considered in Fig.~\ref{mono_Shiba_Sx_magx}, we find that the states with opposite energy also have opposite spin for spin-triplet pairing. We thus conclude that the peculiar occurrence of having the same spin for subgap states with opposite energies is a distinguishing characteristic of a spin-triplet pairing state ($p_x$, $p_y$, $p+ip\,'$ and $f$-wave) and could be used as an experimental signature to identify a spin-triplet order parameter. Moreover, the dependence of the SPLDOS with the direction of the impurity spin is also a characteristic unique to spin-triplet pairing, which additionally could be used to distinguish between a spin-singlet and spin-triplet triplet order parameters.

\subsection{Quasi-particle interference}

The plots above describe the dependence of the average LDOS change induced by an impurity as a function of energy and impurity strength, and thus tell us at which energy the subgap states form. In what follows we are interested in the spatial dependence of the subgap states. In particular, we study the Fourier transform of the LDOS and of the SPLDOS at a given subgap energy peak as a function of momentum. We primarily focus on two different peak energies, $E=0$ and $E \ne 0$. In Tables~\ref{Table2}  and \ref{Table3} we provide the values of the scalar impurity strength $U$, and of the magnetic impurity strength $J$, and the corresponding peak energies for each order parameter symmetry. All energies are given in units of $t$, i.e.\ we set $t=1$. 
\begin{table}
\renewcommand*\arraystretch{1.4}
\begin{center}
\begin{tabular}{c|c|c|c}
\multicolumn{1}{c|}{Energy} & \multicolumn{1}{c|}{Scalar} & \multicolumn{1}{c|}{Magnetic} & \multicolumn{1}{c}{Symmetry} \\
\hline
0.02 & & J=3.5 & $s_{ext}$\\
\hline
0.1 & & J=2.5 & $d_{x^2-y^2}$\\
\hline
0.1 & U=2.5 & J=2.5 & $d_{xy}$\\
\hline
0.1 & U=3 & J=3 & $d_{x^2-y^2} + i d_{xy}$\\
\hline
0.1 & U=1.5 & J=1.5 & $p_x$\\
\hline
0.1 & U=1.5 & J=1.5 & $p_y$\\
\hline
0.1 & U=2 & J=2 & $p_x + i p_y$\\
\hline
0.2 & U=1.5 & J=3 &  $f_{x(x^2-y^2)} $ \\
\hline
0.2 & U=3 & J=1.5 & $s_{\rm ON}$
\end{tabular}
\captionof{table}{Values of impurity strength and energy used to generate the QPI plot for scalar ($U$) and magnetic ($J$) impurities for all considered order parameter symmetry  in Fig.~\ref{mono_QPI_scalimp_e_en0}, }
\label{Table2}
\end{center}
\end{table}
\begin{table}
\renewcommand*\arraystretch{1.4}
\begin{center}
\begin{tabular}{c|c|c|c}
\multicolumn{1}{c|}{Energy} & \multicolumn{1}{c|}{Scalar} & \multicolumn{1}{c|}{Magnetic} & \multicolumn{1}{c}{Symmetry} \\
\hline
0 &  & J=2 & $s_{ext}$\\
\hline
0 &  & J=5 & $d_{x^2-y^2}$\\
\hline
0 & U=5 & J=5 & $d_{xy}$\\
\hline
0 & U=6 & J=6 & $d_{x^2-y^2} + i d_{xy}$\\
\hline
0 & U=5 & J=5 & $p_x$\\
\hline
0 & U=5 & J=5 & $p_y$\\
\hline
0 & U=6 & J=6 & $p_x + i p_y$\\
\hline
0 & U=6 & J=6 &  $f_{x(x^2-y^2)} $ \\
\hline
0 & U=2.5 & J=2.5 & $s_{\rm ON}$
\end{tabular}
\captionof{table}{
Values of impurity strength corresponding to a zero-energy subgap state used to generate the QPI for scalar ($U$) and magnetic ($J$) impurities for all considered order parameter symmetry in Figs.~\ref{mono_QPI_scalimp_e_e0},~\ref{mono_QPI_scalimp_e_e0s},\ref{mono_QPI_sz_magz_e_e0}, and~\ref{mono_QPI_sz_magx_e_e0}.}
\label{Table3}
\end{center}
\end{table}

In the following we also only plot the absolute value of $\delta \rho ( \protect \textbf{q})$ and of $\delta S_{\alpha} ( \protect \textbf{q})$, $\alpha=x,y,z$. This is because, to the hexagonal structure of the lattice, these are generally complex quantities that have both non-zero real and imaginary parts. However, at present it is very hard to distinguish experimentally between their real and imaginary parts: experiments calculating the QPI patterns based on fast Fourier transform (FFT) cannot keep track precisely neither of the phase, nor of the sign. Moreover, spin-polarized STM experiments with corresponding QPI are still in their infancy. Thus, in order to avoid the overload of information, we only focus on the absolute Fourier transform values. If more accurate experimental data become available this study can easily be extended and refined to take into account separately the real and imaginary parts of the Fourier transforms of both the LDOS and SPLDOS.

 \begin{figure*}[!htb]
\begin{center}
\resizebox{1.6\columnwidth}{!}{%
    \begin{tabular}{ c  c  c  c}
      \thead{\includegraphics[width=4cm]{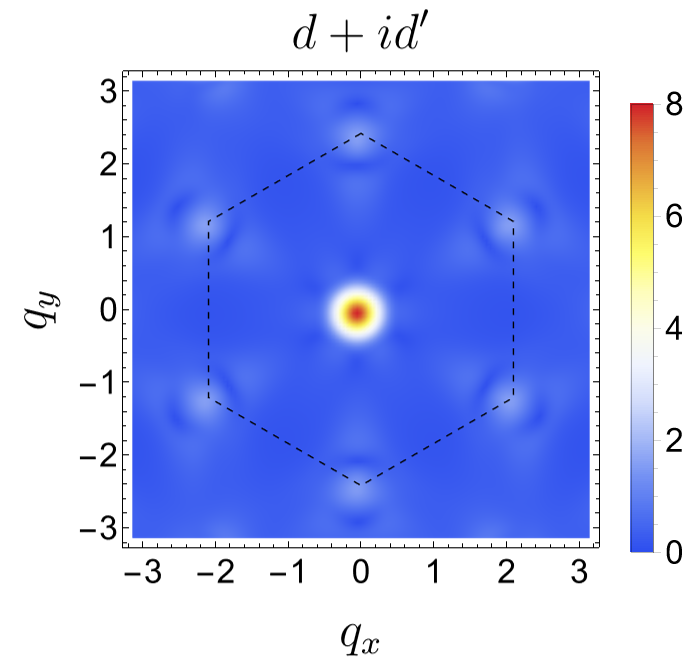}} 
   & \thead{\includegraphics[width=4.2cm]{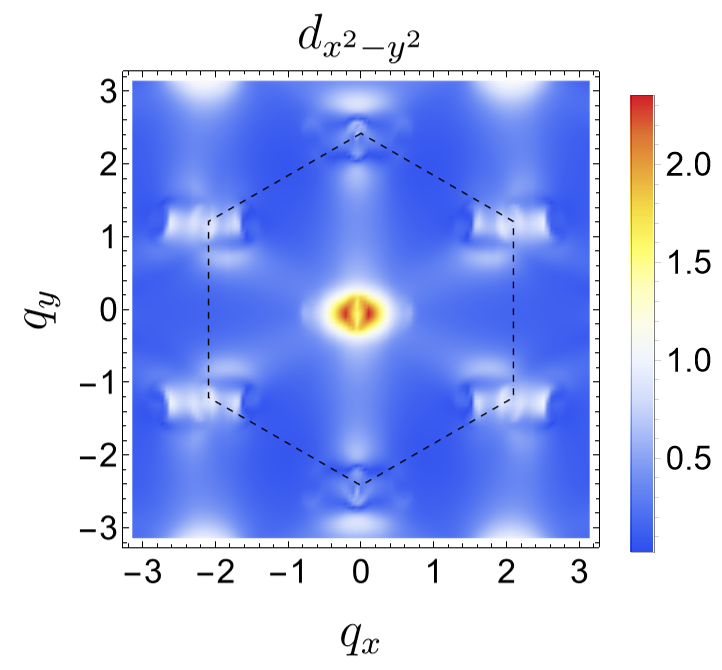}}
   & \thead{\includegraphics[width=4.2cm]{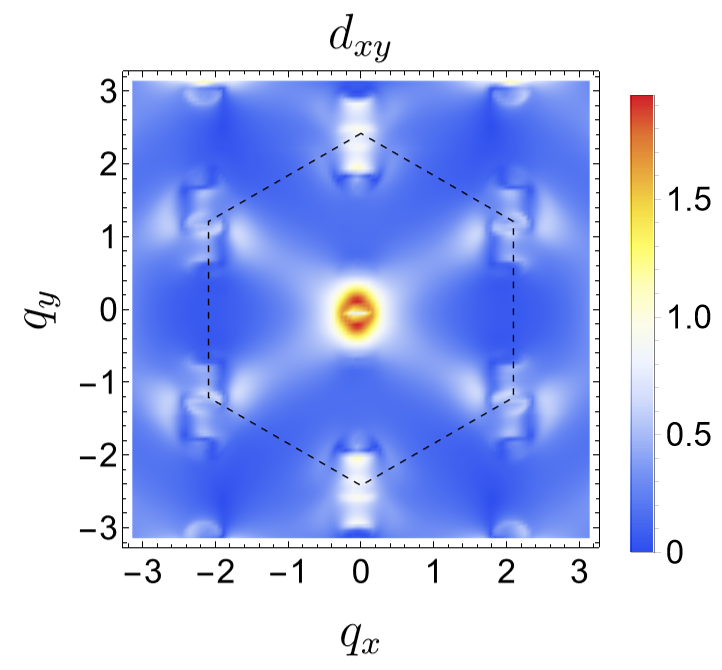}} 
       \vspace{-0.4cm}
   \\
      \thead{\includegraphics[width=4cm]{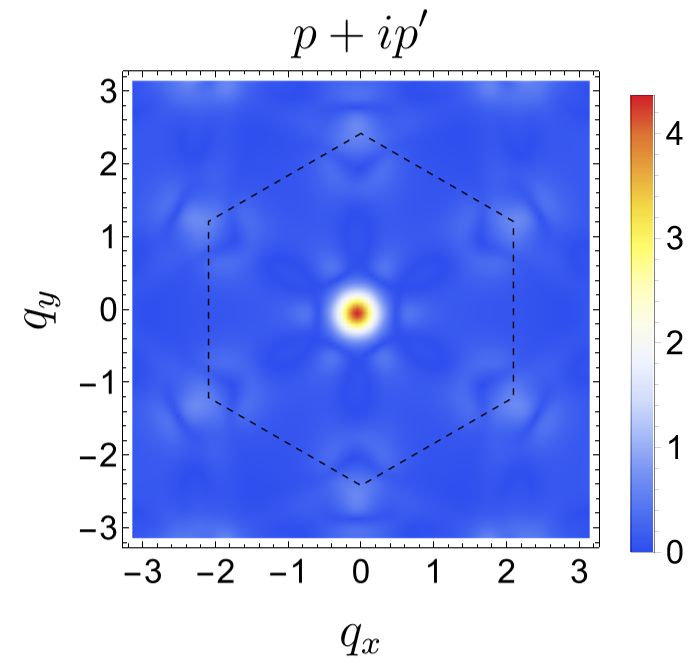} } 
      &  \thead{\includegraphics[width=4.2cm]{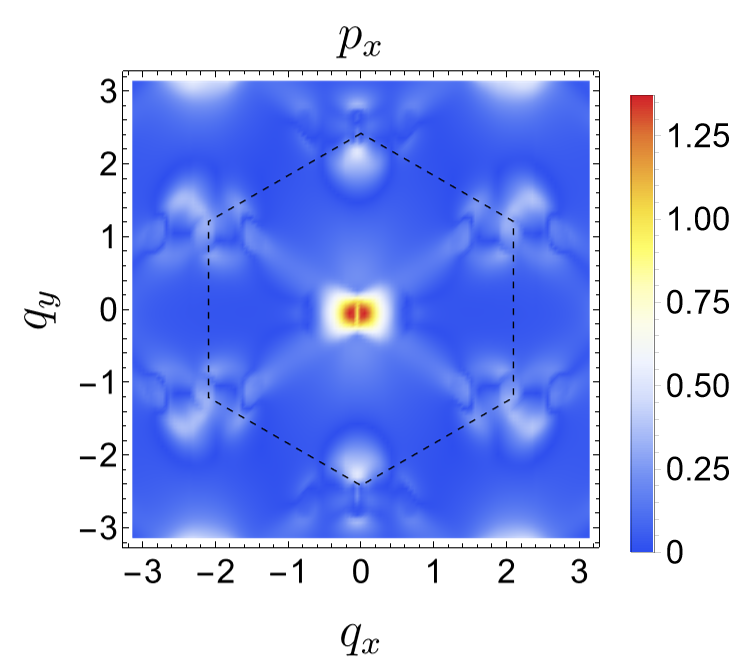}}  
      & \thead{ \includegraphics[width=4.2cm]{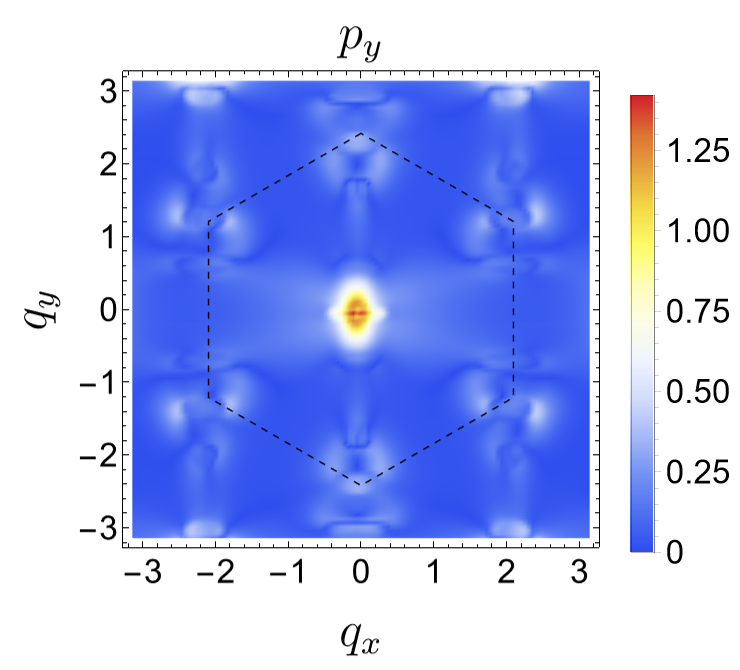} }
      %\thead{ \includegraphics[width=4cm]{QPI_on_scal__LDOS.png}}
      & \thead{\includegraphics[width=4.1cm]{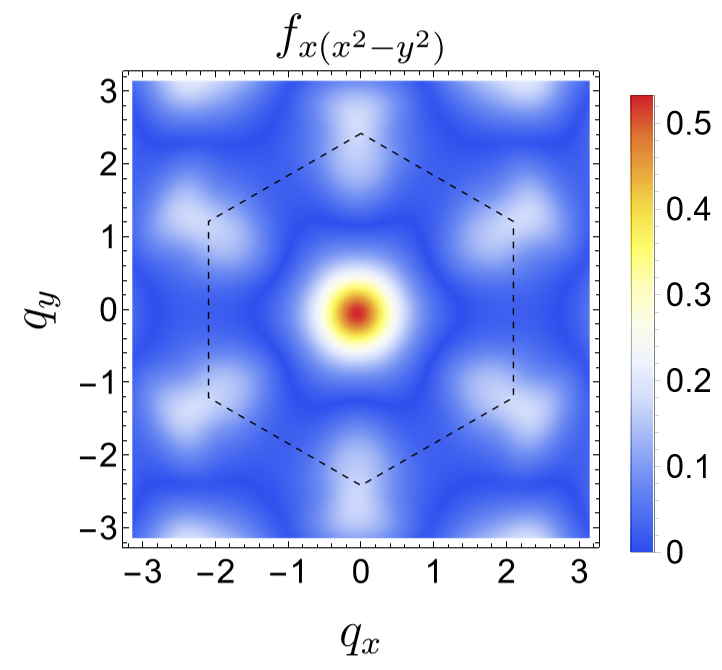} }
       %\thead{\includegraphics[width=4cm]{QPI_s_{ext}_scal__LDOS.png}}
       \end{tabular}
    }%
        \vspace{-0.7cm}
  \end{center}
  \caption{ $|\delta \rho ( \protect \textbf{q})|$ at the values of  impurity strength $U$ and energy provided in Table~\ref{Table2} for a scalar impurity. We take $\mu = 0.4 t$ and $\Delta_0 = 0.4t$. The Brillouin zone is indicated by dashed lines.}
  \label{mono_QPI_scalimp_e_en0}
 \end{figure*}

    \begin{figure*}[!htb]
\begin{center}
\resizebox{1.6\columnwidth}{!}{%
    \begin{tabular}{ c  c  c  c}
      \thead{\includegraphics[width=4cm]{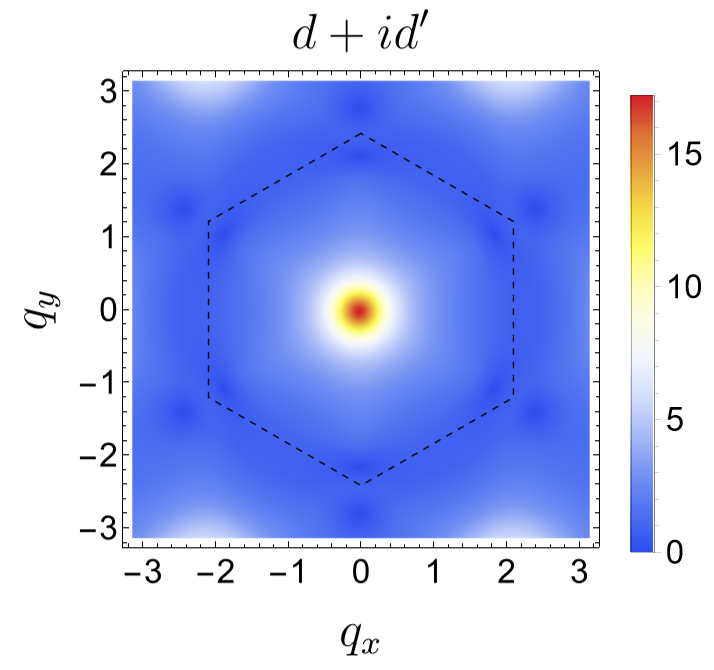}} 
   & \thead{\includegraphics[width=4.2cm]{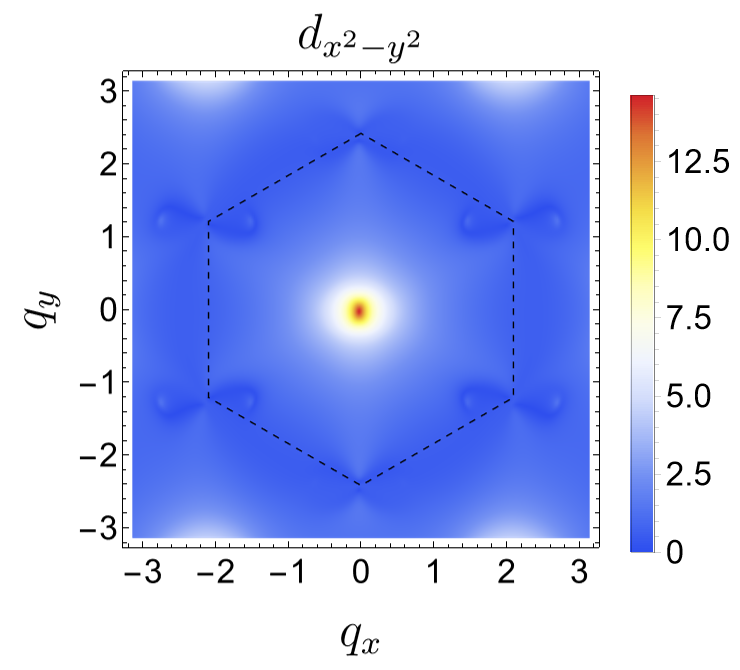}}
   & \thead{\includegraphics[width=4cm]{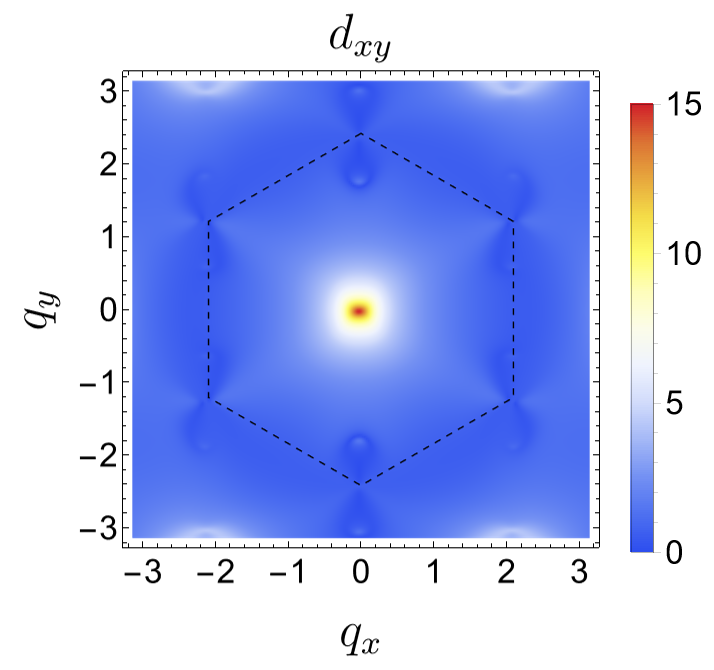}} 
       \vspace{-0.6cm}
   \\
      \thead{\includegraphics[width=4cm]{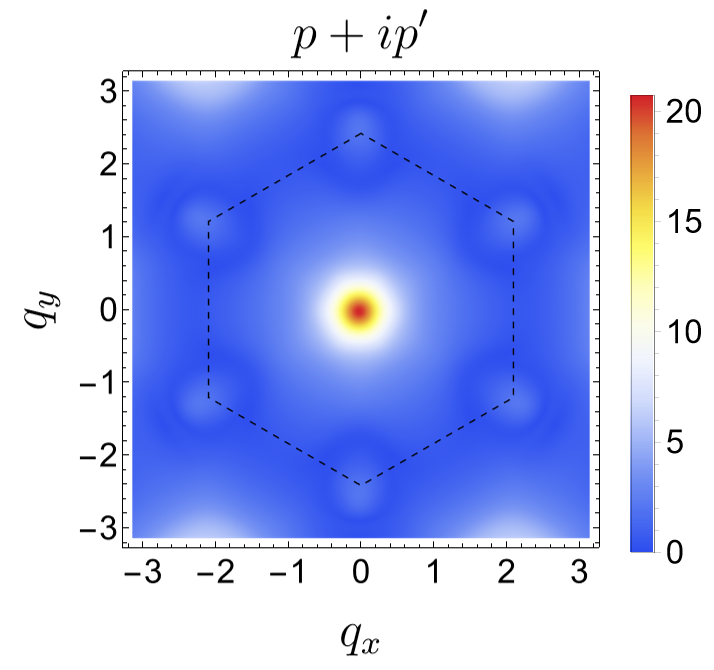} } 
      &  \thead{\includegraphics[width=4cm]{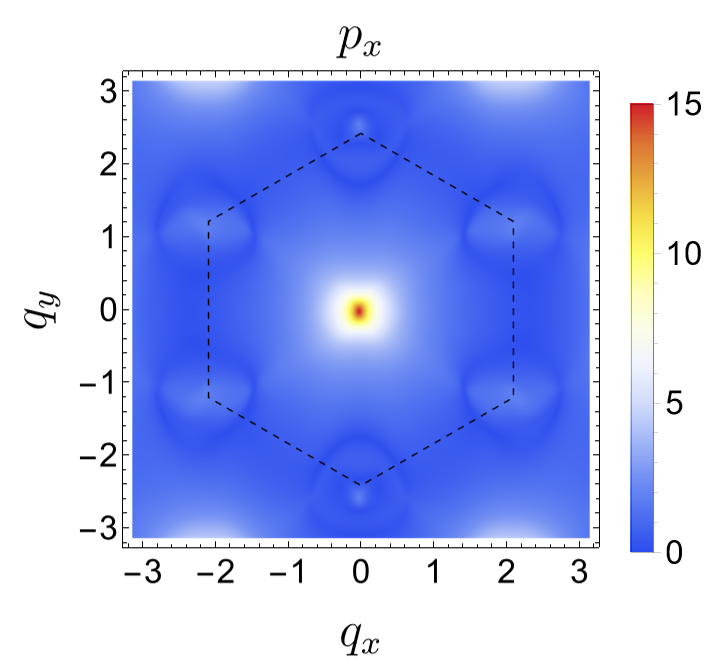}}
      & \thead{ \includegraphics[width=4.2cm]{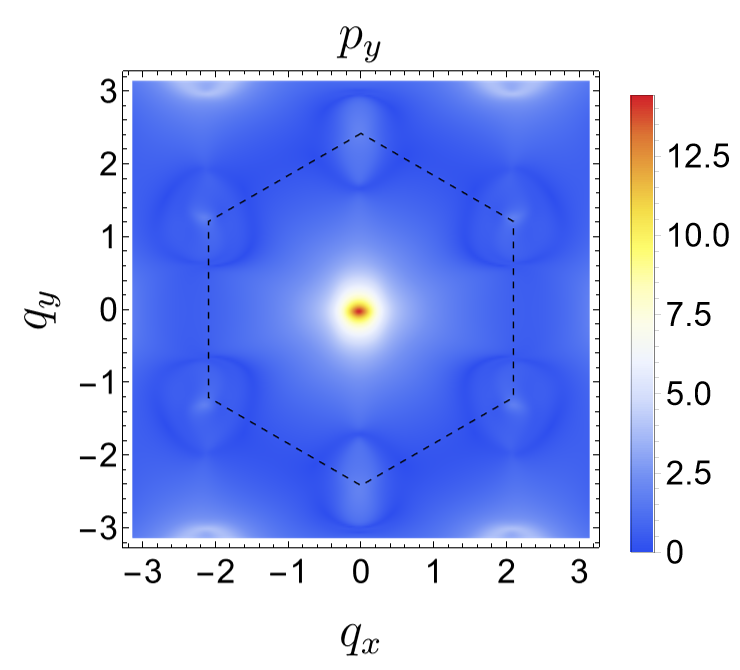} }
      %\thead{ \includegraphics[width=4cm]{QPI_e0_on_scal__LDOS.png}}
      & \thead{\includegraphics[width=4.1cm]{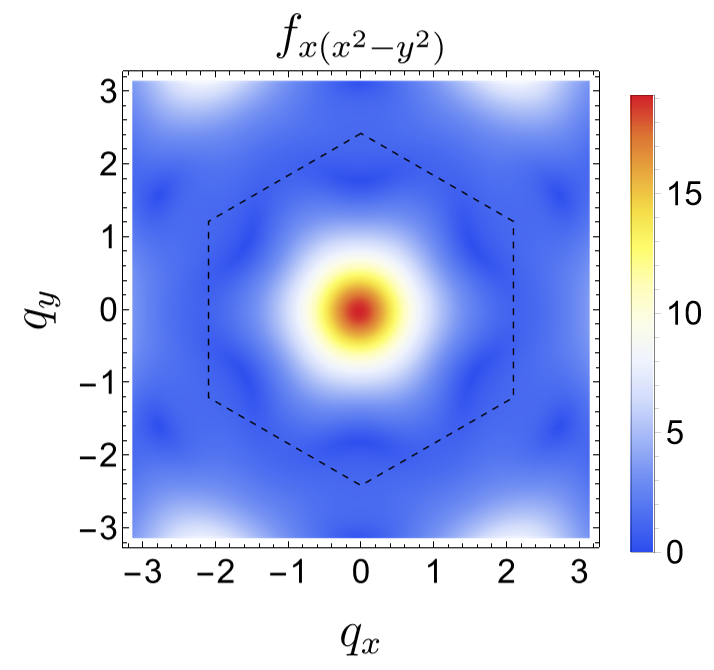} }
       %\thead{\includegraphics[width=4cm]{QPI_e0_s_{ext}_scal__LDOS.png}}
    \end{tabular}
    }%
        \vspace{-0.7cm}
  \end{center}
  \caption{ $|\delta \rho ( \protect \textbf{q})|$ at zero energy and the corresponding impurity strength values $U$ in Table~\ref{Table3} for a scalar impurity.  We take $\mu = 0.4 t$ and $\Delta_0 = 0.4t$. The Brillouin zone is indicated by dashed lines.}
  \label{mono_QPI_scalimp_e_e0}
 \end{figure*}
 
Similarly to above, we start by considering a scalar impurity ($U \ne 0$ and $J=0$), and calculate the QPI patterns for both a non-zero energy in Fig.~\ref{mono_QPI_scalimp_e_en0}, and at zero energy in Fig.~\ref{mono_QPI_scalimp_e_e0}. Since the 
 $s_{\rm ON}$-wave and $s_{\rm ext}$-wave order parameters do not exhibit any subgap states for a scalar impurity we do not include them in the scalar-impurity QPI analysis. Overall, the QPI patterns are dominated by a central feature at the center of the Brillouin zone ($\Gamma$-point), corresponding to intra-nodal scattering (in the normal-state band structure) of the electrons by the impurity, and by six features localized at the corners of the Brillouin zone ($K$-points) corresponding to inter-nodal scattering.

We further note that the gapless nodal states clearly produce a QPI pattern that breaks the six-fold symmetry, while the gapped states, i.e the $f$-wave, and the chiral $d+id\,'$- and $p+ip\,'$-wave states, all show QPI patterns that preserve the full rotation symmetry of the lattice. This is fully consistent with the symmetries of the SC order parameter, modulo the order parameter phase that might change sign, but which seemingly does not affect the QPI in contrast to the case of $d$-wave cuprates \cite{pereg2008magnetic}. It is also fully consistent with the fact that these states have a symmetry-preserving SC band structure \cite{pangburn2022superconductivity} for all gapped states. QPI would thus be a good experimental tool to distinguish between nodal states that break rotation symmetry and the gapped states which do not.  A similar observation has already been made in Refs.~\onlinecite{lothman2014defects,Awoga2018}, when comparing the nodal $d$-wave states and the chiral $d$-wave states. Here we establish that this also holds for both spin-singlet $d$-wave and spin-triplet $p$-wave symmetries in graphene.

\begin{figure}
\begin{center}
\resizebox{0.8\columnwidth}{!}{%
    \begin{tabular}{ c  c  c }
      \thead{ \includegraphics[width=4cm]{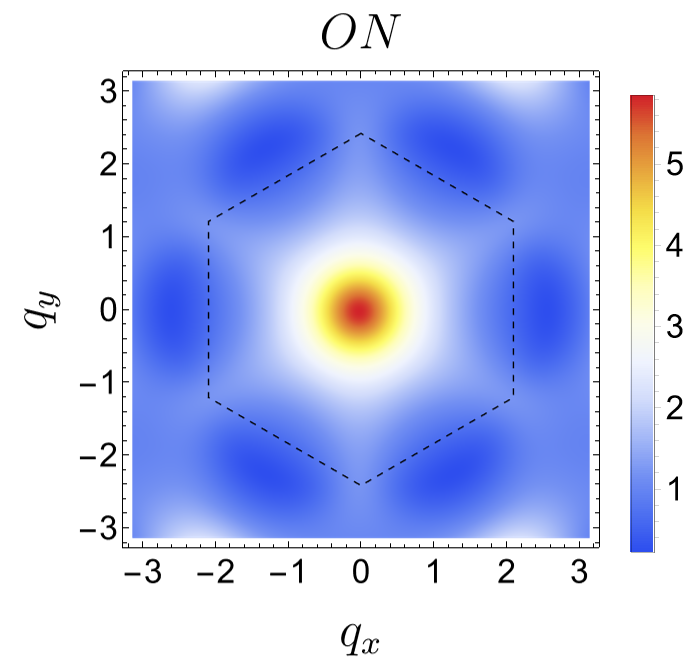}}
      & \thead{\includegraphics[width=4cm]{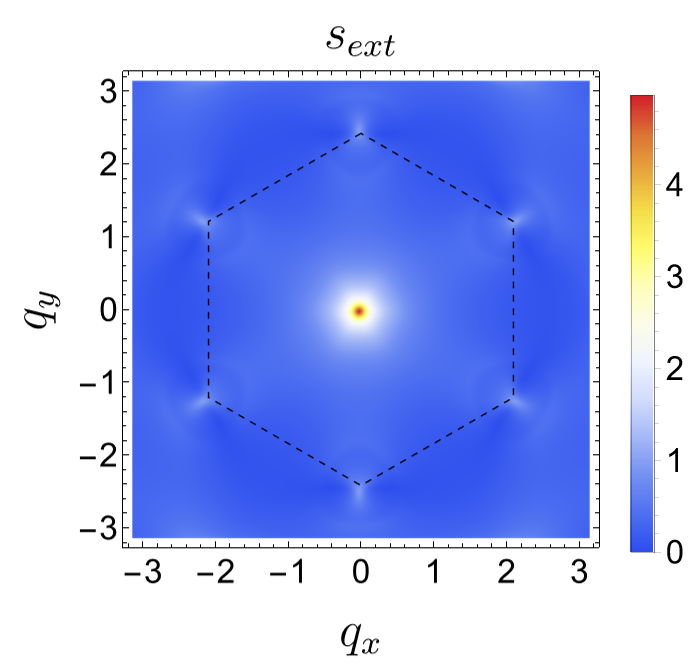}}
    \end{tabular}
    }%
        \vspace{-0.7cm}
  \end{center}
  \caption{ $|\delta \rho ( \protect \textbf{q})|$ at zero energy and the corresponding impurity strength $J_z=J$ values in Table~\ref{Table3}. We take $\mu = 0.4 t$ and $\Delta_0 = 0.4t$. The Brillouin zone is indicated by dashed lines.}
  \label{mono_QPI_scalimp_e_e0s}
 \end{figure}

We next analyze the QPI patterns generated by a magnetic impurity  ($J \ne 0$ and $U=0$).  We here only plot the QPI corresponding to the zero-energy peaks, as we find that the effect for the non-zero energy subgap states is very similar. We further find that in the presence of a $z$-magnetic impurity we recover the same features for $|\delta \rho ( \protect \textbf{q})|$ as those for a scalar impurity depicted in Fig.~\ref{mono_QPI_scalimp_e_e0}. The main difference is that, in the presence of a magnetic impurity the states with $s_{\rm ON}$-wave and $s_{\rm ext}$-wave order parameters also exhibit subgap states. Thus, to avoid repetition, in Fig.~\ref{mono_QPI_scalimp_e_e0s} we only plot $|\delta \rho ( \protect \textbf{q})|$ for these two $s$-wave states, for which the corresponding values of the $z$-magnetic impurity strength, $J=J_z$, are given in Table \ref{Table3}. 

\begin{figure*}[!htb]
\begin{center}
%\resizebox{2\columnwidth}{!}{
\resizebox{2\columnwidth}{!}{%
    \begin{tabular}{ c  c  c c c}
 \thead{ \includegraphics[width=5cm]{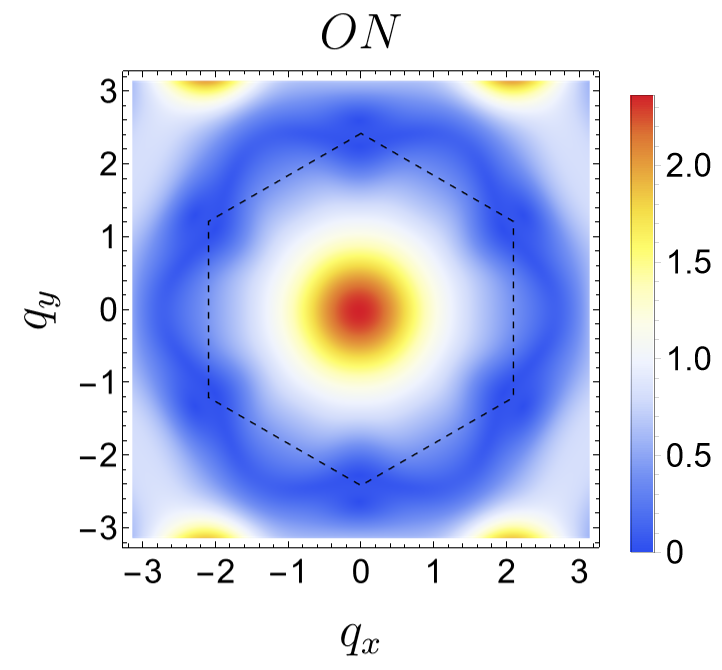}}
 & \thead{\includegraphics[width=5cm]{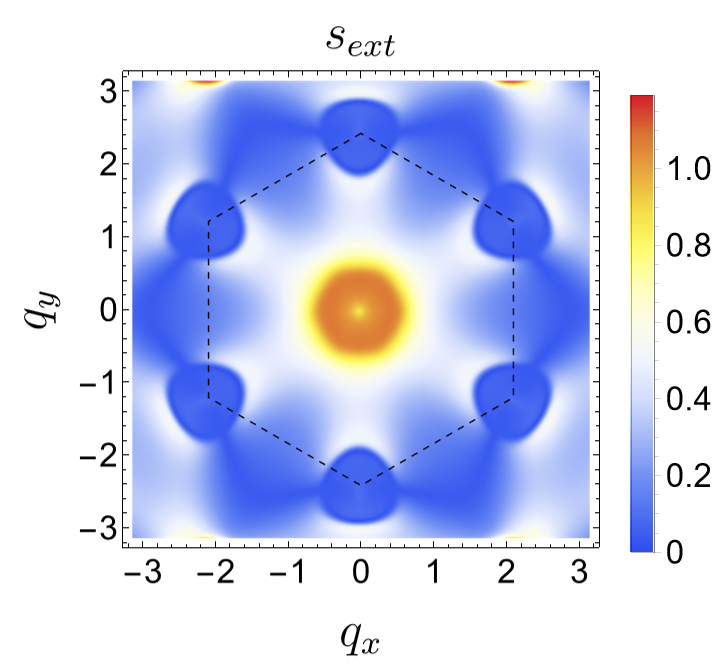}}
    &  \thead{\includegraphics[width=5cm]{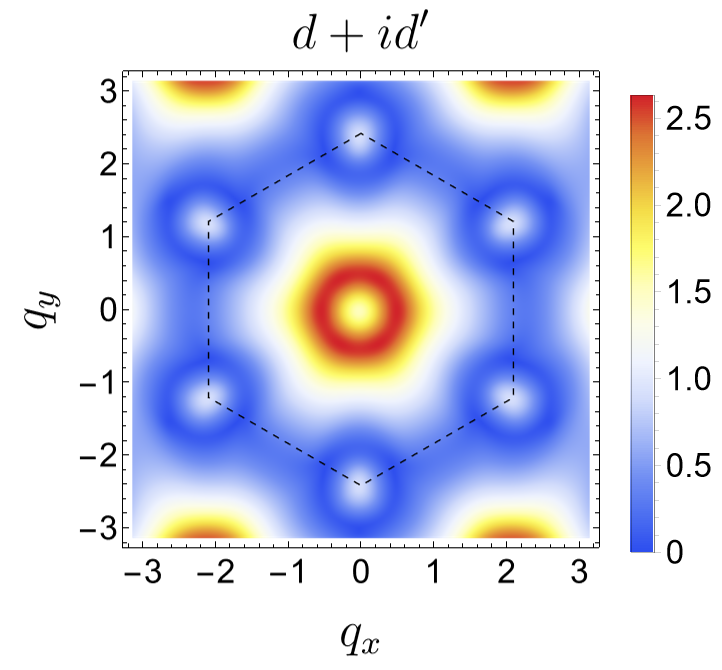}} 
   & \thead{\includegraphics[width=5cm]{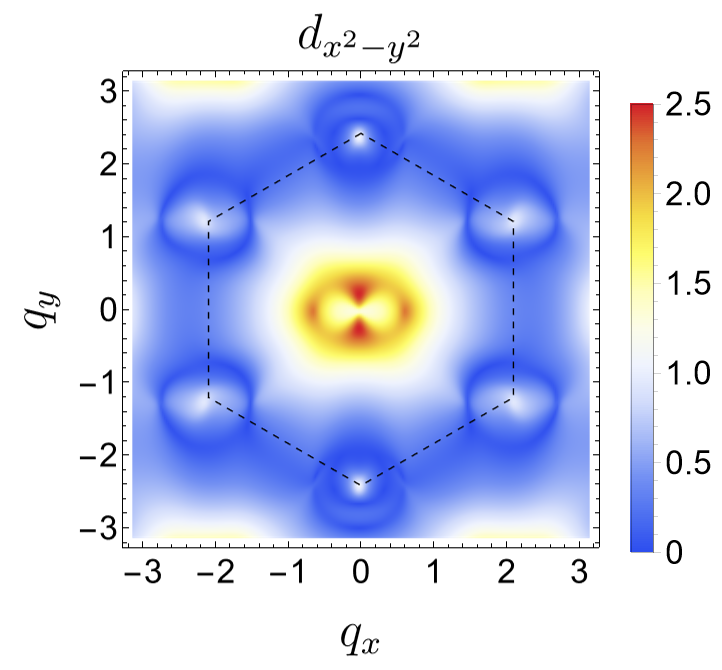}}
   & \thead{\includegraphics[width=5cm]{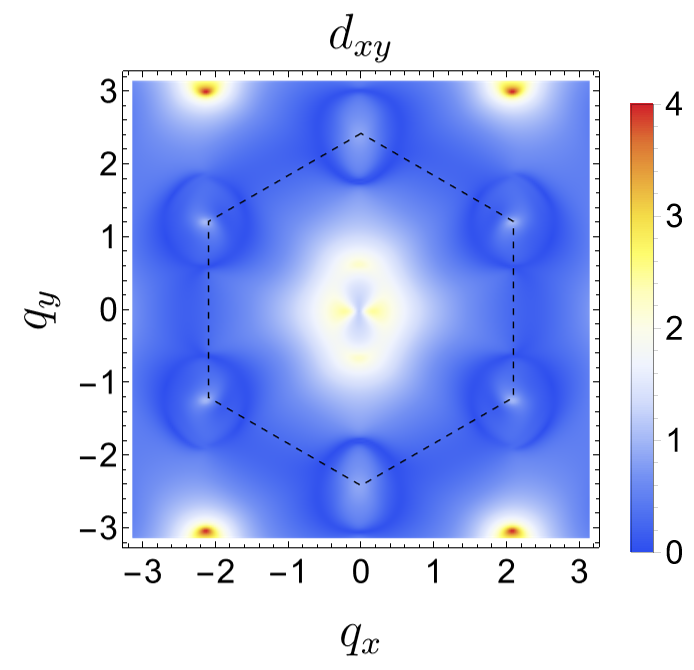}} 
       \vspace{-0.4cm}
   \\
      \thead{\includegraphics[width=5cm]{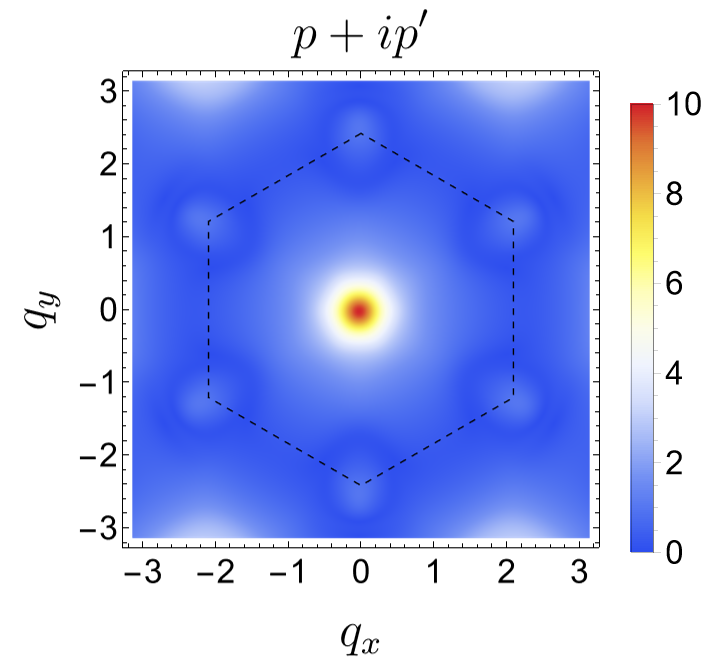} } 
      &  \thead{\includegraphics[width=5cm]{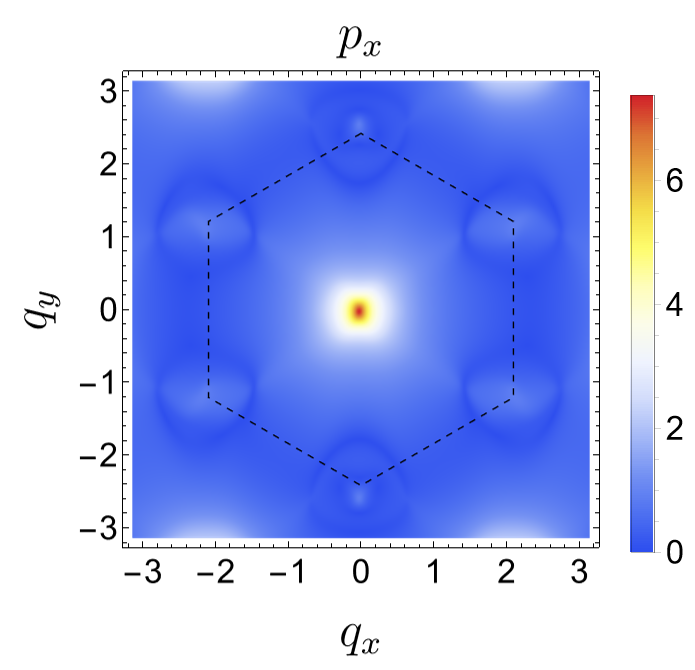}}
      & \thead{ \includegraphics[width=5cm]{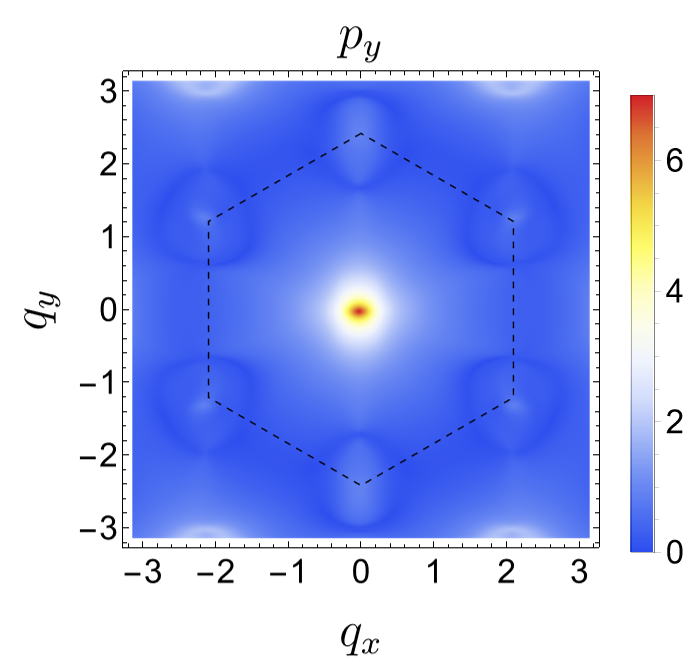} } 
      & \thead{\includegraphics[width=5cm]{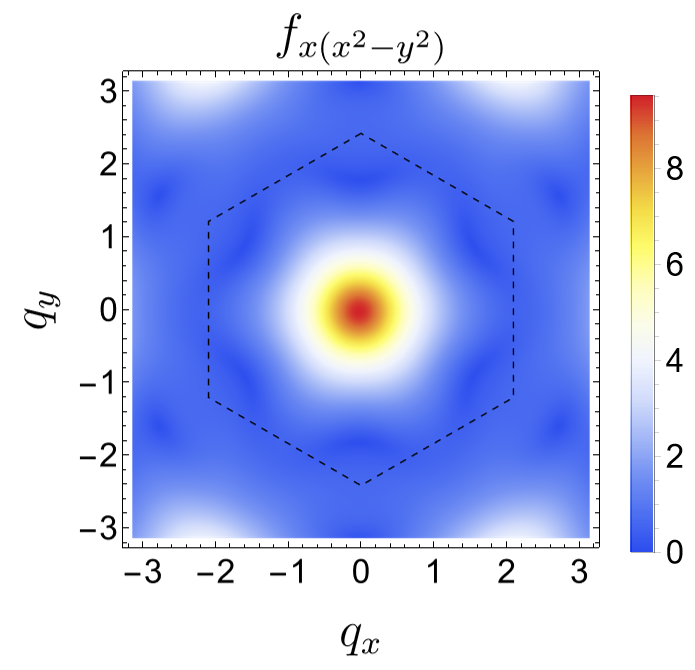} }
    \end{tabular}
    }%
        \vspace{-0.7cm}
  \end{center}
  \caption{$|\delta S_z ( \protect \textbf{q})|$ at zero energy and corresponding impurity strength values $J_z=J$ in Table~\ref{Table3}. We take $\mu = 0.4 t$ and $\Delta_0 = 0.4t$. The Brillouin zone is indicated by dashed lines.}
  \label{mono_QPI_sz_magz_e_e0}
 \end{figure*}

\begin{figure}[!htb]
\begin{center}
\resizebox{0.8\columnwidth}{!}{%
    \begin{tabular}{ c  c  c }
      \thead{\includegraphics[width=4cm]{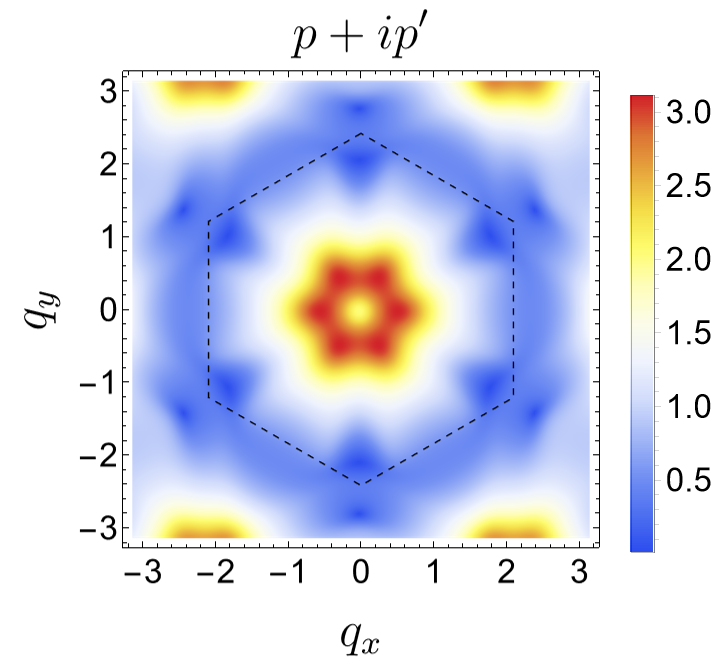}} 
   & \thead{\includegraphics[width=4cm]{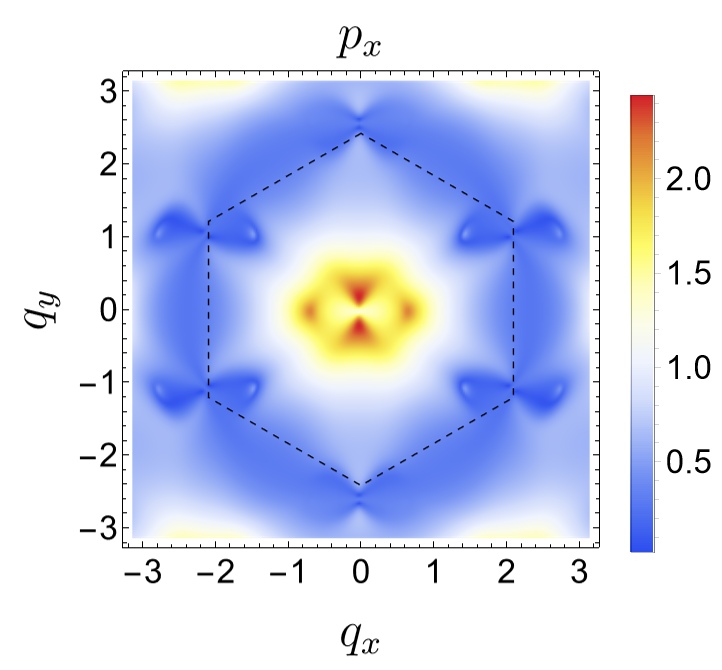}}
       \vspace{-0.4cm}
   \\
    \thead{\includegraphics[width=4cm]{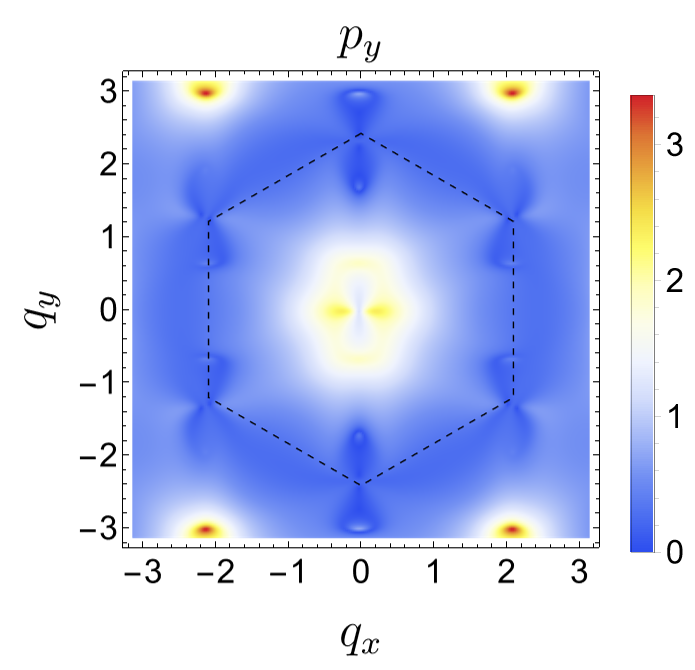}} 
&      \thead{\includegraphics[width=4cm]{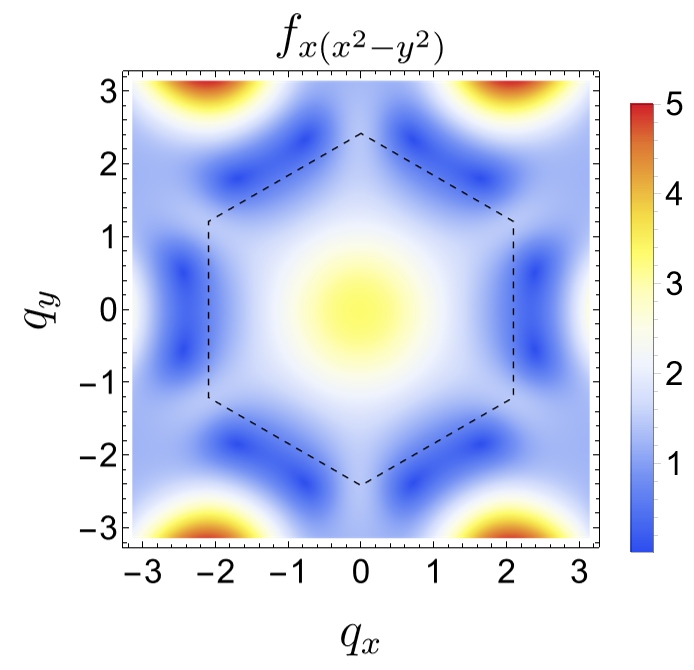} } 
    \end{tabular}
    }%
        \vspace{-0.7cm}
  \end{center}
  \caption{ $|\delta S_x(\textbf{q})|$ at zero energy and corresponding impurity strength values $J_x=J$ in Table~\ref{Table3} for a $x$-magnetic impurity. We take $\mu = 0.4 t$ and $\Delta_0 = 0.4t$. The Brillouin zone is indicated by dashed lines.}
  \label{mono_QPI_sz_magx_e_e0}
 \end{figure}

While $|\delta \rho ( \protect \textbf{q})|$ does not show any significant differences, $|\delta S_z ( \protect \textbf{q})|$, i.e.\ the spin-polarized LDOS, shows more interesting features, which we plot in Fig.~\ref{mono_QPI_sz_magz_e_e0} for the same $z$-magnetic impurity. The main differences from the QPI of the scalar impurity are a ring-like feature arising in the center of the Brillouin zone in the $d$-wave SC states, as well as a reduction in the asymmetry for the $K$-points features.\\ 
We also note that for the spin-triplet pairing states, the $x$-SPLDOS QPI for a $x$-magnetic impurity is different from the $y$ and $z$ ones and becomes more reminiscent of the spin-singlet ones, where there is no dependence on $x,y,z$-magnetic impurity directions.
For example, as depicted in Fig.~\ref{mono_QPI_sz_magx_e_e0}, the spin-polarized QPI's for $p_x$/$p_y$/$p+ip\,'$-wave states at zero energy and the corresponding impurity strength values $J_x=J$ provided in Table \ref{Table3}, acquire more similarities  to the $d_{xy}$/$d_{x^2-y^2}$/$d+id\,'$-wave states, in that they show a ring of high intensity for the feature in the center of the Brillouin zone and an increased asymmetry for the features at the corners of the Brillouin zone.

\vspace{5in}

\section{Multilayer graphene\label{sec:Multilayer}}
We next consider both AB-stacked bilayer graphene and ABC or ABA-stacked trilayer graphene. We first note that for most of the order parameter symmetries, the number of subgap states, as well as their impurity-strength dependence and their spin dependences are quite universal, generic features, and do not depend on the number of layers or the stacking. In what follows, to avoid redundancy, we only present the LDOS and SPLDOS results when there is a difference from the generic case. In particular, we find differences for ABC-stacked trilayer graphene in the presence of gapless $d_{xy}$-, $d_{x^2-y^2}$-, $p_x$-, or $p_y$-wave order parameters. 

\begin{figure}[!htb]
\begin{center}
\resizebox{0.8\columnwidth}{!}{%
    \begin{tabular}{ c  c  c }
   %  \thead{
  % \includegraphics[width=4cm]{ABA_pho_didg1_02_g3_0.png}} 
   %& \thead{
   %\includegraphics[width=4cm]{ABA_pho_dx2y2g1_02_g3_0.png}} 
   %& \thead{
  % \includegraphics[width=4cm]{ABA_pho_dxyg1_02_g3_0.png}}
   %
%   \\
   %   \thead{\includegraphics[width=3.5cm]{ABA_pho_pipg1_02_g3_0.png} } 
      %& \thead{
   %\includegraphics[width=4cm]{ABA_pho_pxg1_02_g3_0.png}} 
   %& \thead{ 
   %\includegraphics[width=4cm]{ABA_pho_pyg1_02_g3_0.png}}  
   %   
      %\\
     %\thead{
   %\includegraphics[width=4cm]{ABA_pho_ONg1_02_g3_0.png}} 
  % & \thead{
   %\includegraphics[width=4cm]{ABA_pho_fxg1_02_g3_0.png}} 
  % & \thead{
   %\includegraphics[width=4cm]{ABA_pho_sg1_02_g3_0.png}}
%\\
   %   \thead{
   %\includegraphics[width=4cm]{ABC_pho_didg1_02_g3_0.png}} 
    \thead{
   \includegraphics[width=4cm]{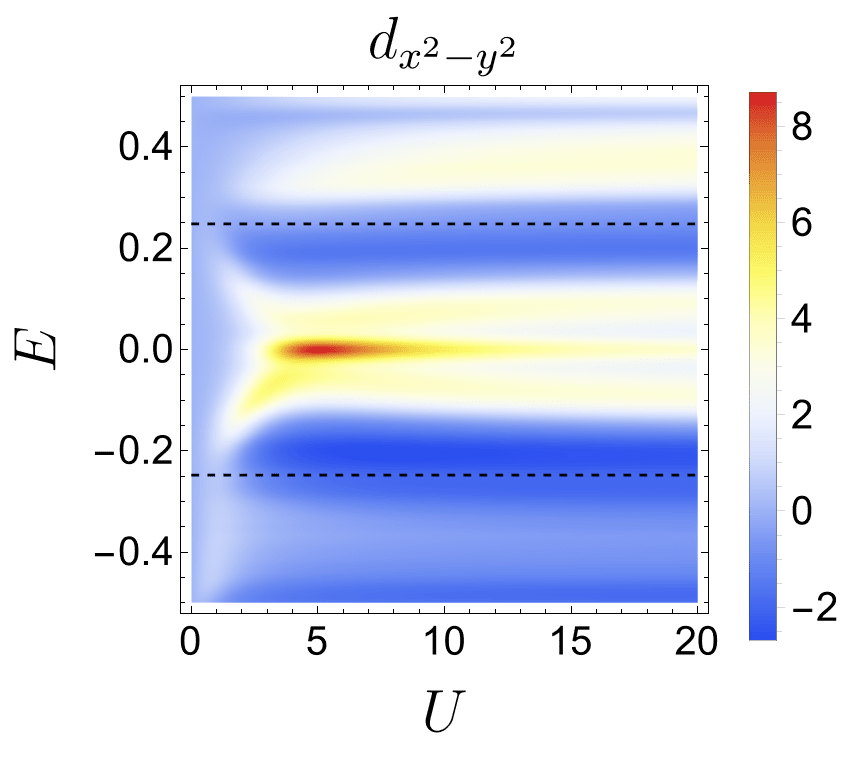}} 
   & \thead{
   \includegraphics[width=4cm]{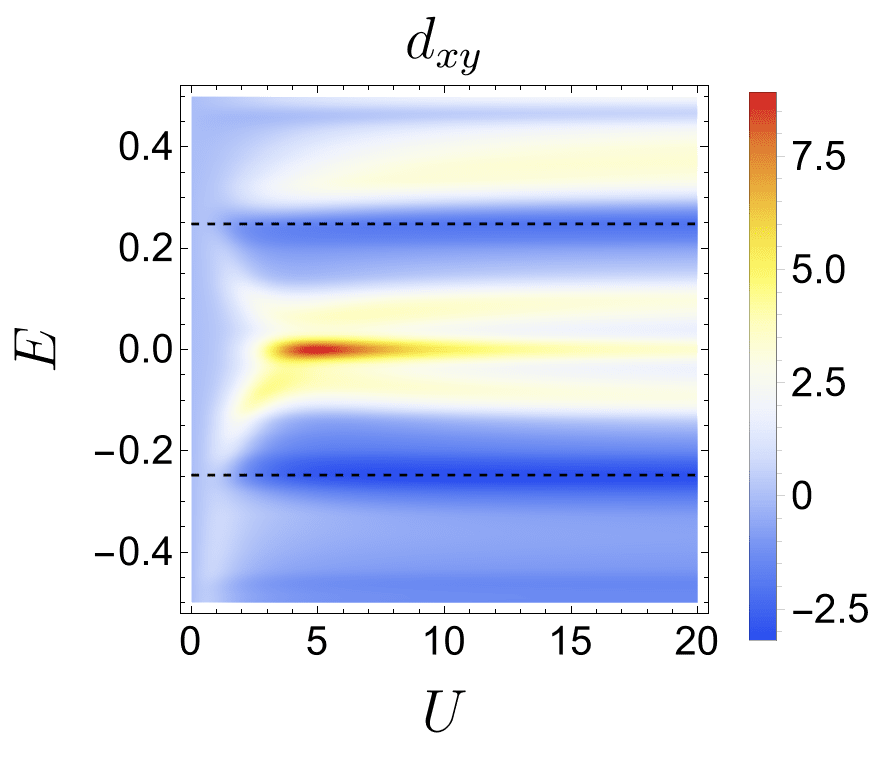}}  
       \vspace{-0.4cm}
   \\
%      \thead{\includegraphics[width=4cm]{ABC_pho_pipg1_02_g3_0.png} } 
      %&
      \thead{
   \includegraphics[width=4cm]{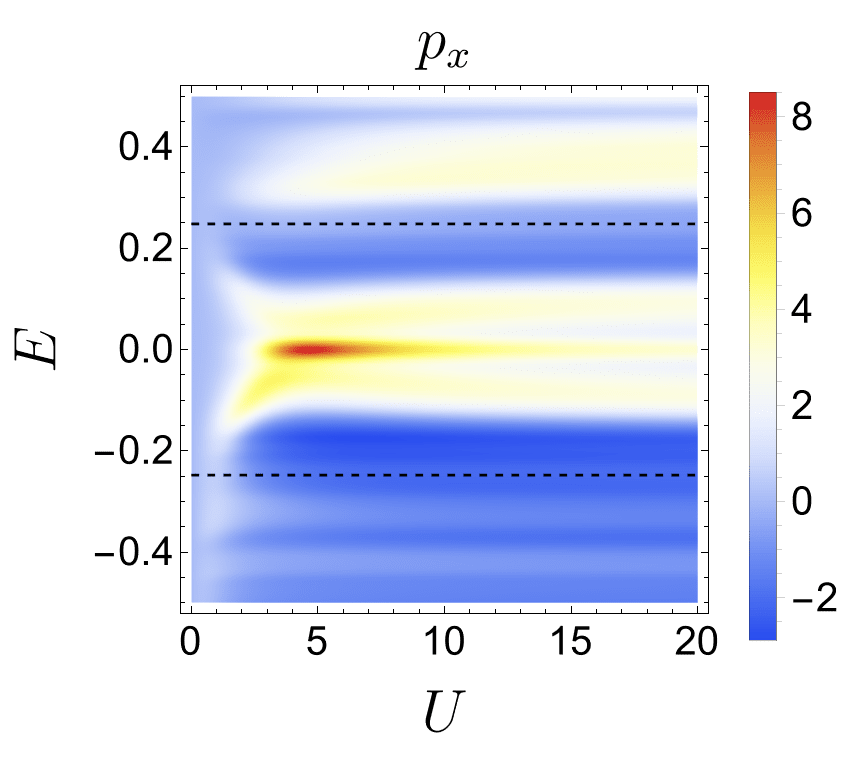}} 
   & \thead{
   \includegraphics[width=4cm]{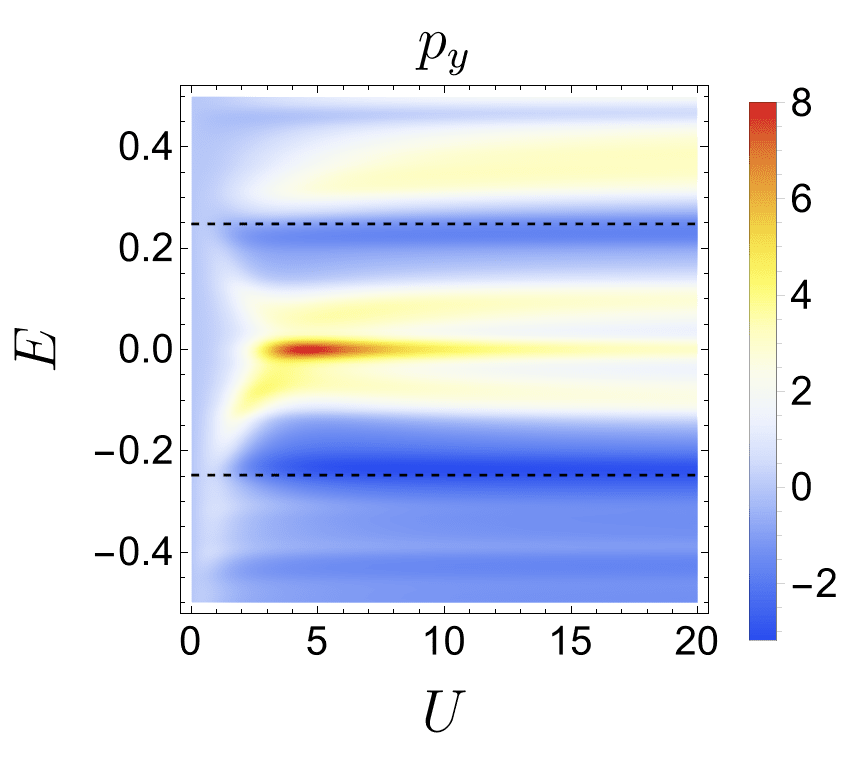}} 
      %\\
  %\thead{
   %\includegraphics[width=4cm]{ABC_pho_ONg1_02_g3_0.png}} 
  % & \thead{
   %\includegraphics[width=4cm]{ABC_pho_fxg1_02_g3_0.png}} 
   %& \thead{
   %\includegraphics[width=4cm]{ABC_pho_sg1_02_g3_0.png}}

    \end{tabular}
    }%
        \vspace{-0.7cm}
  \end{center}
  \caption{$\delta \rho (E)$ as a function of energy and impurity strength $U$ for a scalar impurity at $\Delta_0=0.4$ and $\mu = 0.4$ for the nodal SC states in ABC-stacked trilayer graphene. The dotted lines indicate the gap edge.}
  \label{trilay_Shiba_pho_scal}
 \end{figure}

\begin{figure}[!htb]
\begin{center}
\resizebox{0.8\columnwidth}{!}{%
    \begin{tabular}{ c  c  c }
%      \thead{
 %  \includegraphics[width=4cm]{ABA_pho_mz_didg1_02_g3_0.png}} 
 %  & \thead{
  % \includegraphics[width=4cm]{ABA_pho_mz_dx2y2g1_02_g3_0.png}} 
  % & \thead{
   %\includegraphics[width=4cm]{ABA_pho_mz_dxyg1_02_g3_0.png}}
 %  \\
    %  \thead{\includegraphics[width=4cm]{ABA_pho_mz_pipg1_02_g3_0.png} } 
     % & \thead{ 
  % \includegraphics[width=4cm]{ABA_pho_mz_pxg1_02_g3_0.png}} 
   %& \thead{ 
   %\includegraphics[width=4cm]{ABA_pho_mz_pyg1_02_g3_0.png}}  
    %  
   %   \\
  %\thead{
  % \includegraphics[width=4cm]{ABA_pho_mz_ONg1_02_g3_0.png}} 
%   & \thead{
  % \includegraphics[width=4cm]{ABA_pho_mz_fxg1_02_g3_0.png}} 
  % & \thead{
 %  \includegraphics[width=4cm]{ABA_pho_mz_sg1_02_g3_0.png}}
%\\
   %   \thead{
  %\includegraphics[width=4cm]{ABC_pho_mz_didg1_02_g3_0.png}} 
%   & 
   \thead{
   \includegraphics[width=4cm]{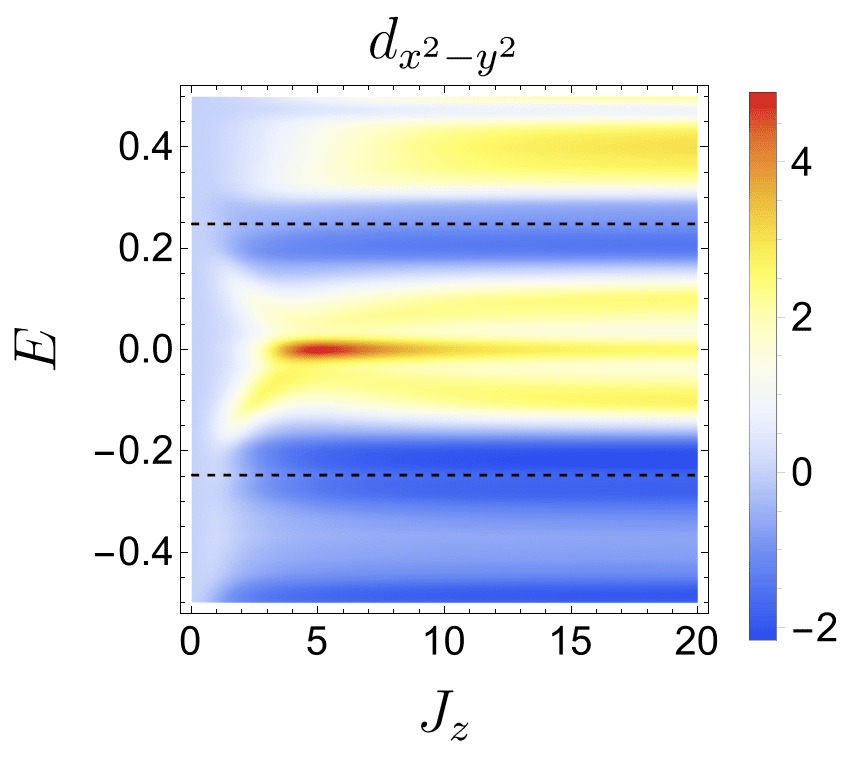}} 
   & \thead{
   \includegraphics[width=4cm]{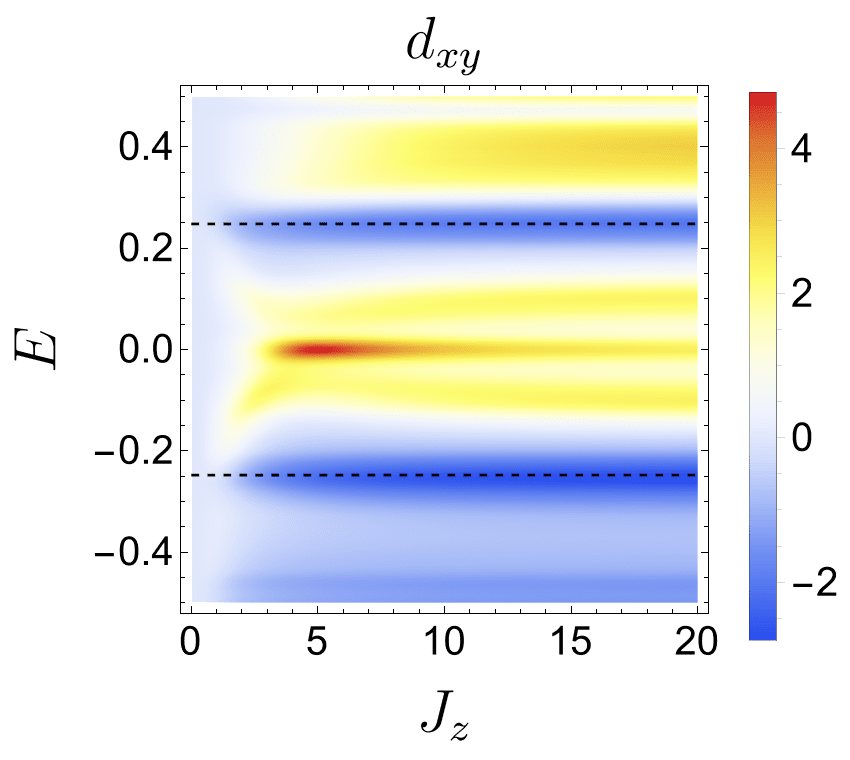}}
       \vspace{-0.4cm}
   \\
    %  \thead{\includegraphics[width=4cm]{ABC_pho_mz_pipg1_02_g3_0.png} } 
    %  & 
      \thead{
   \includegraphics[width=4cm]{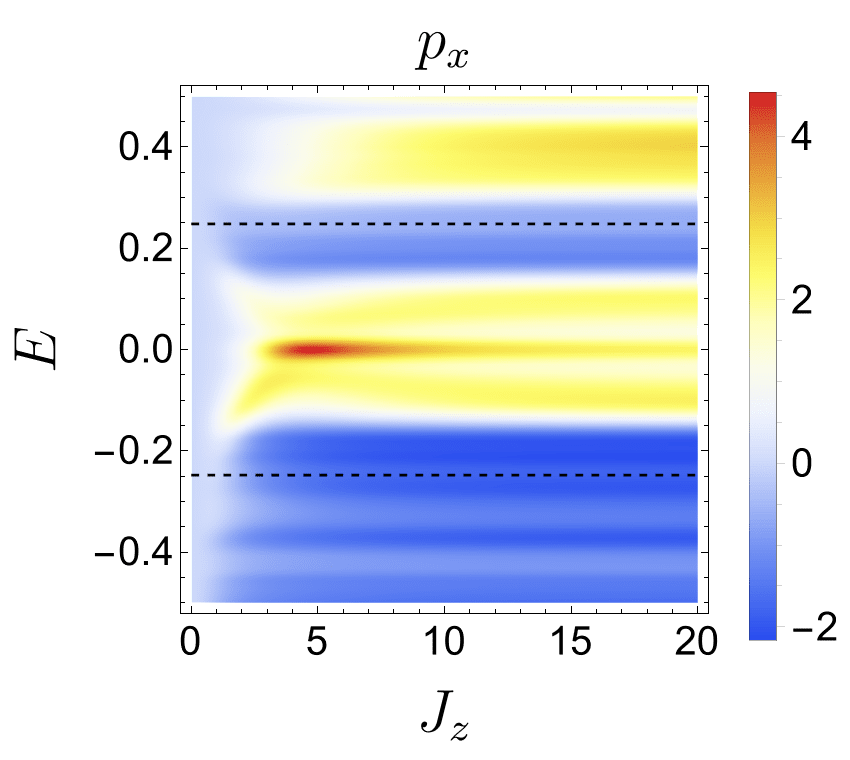}} 
   & \thead{ 
   \includegraphics[width=4cm]{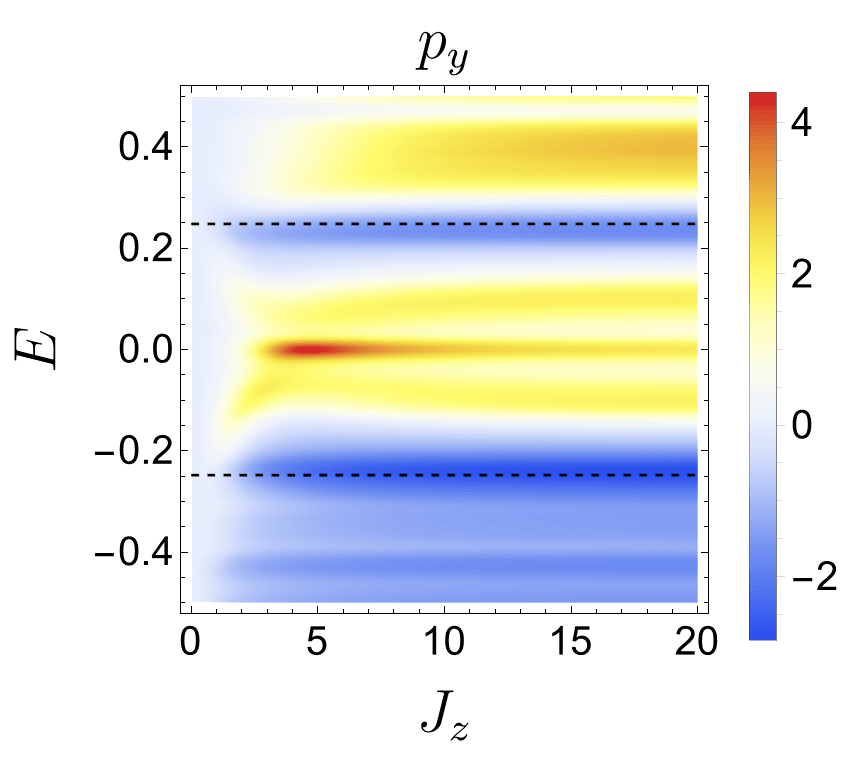}} 
 %     \\
 % \thead{
 %  \includegraphics[width=4cm]{ABC_pho_mz_ONg1_02_g3_0.png}} 
 %  & \thead{
 %  \includegraphics[width=4cm]{ABC_pho_mz_fxg1_02_g3_0.png}} 
%   & \thead{
  % \includegraphics[width=4cm]{ABC_pho_mz_sg1_02_g3_0.png}}

    \end{tabular}
    }%
        \vspace{-0.7cm}
  \end{center}
  \caption{$\delta \rho (E)$ as a function of energy and impurity strength $J_z$ for a $z$-magnetic impurity at $\Delta_0=0.4$ and $\mu = 0.4$ for the nodal SC states in ABC-stacked trilayer graphene. The dotted lines indicate the gap edge.}
  \label{tril_Shiba_mag_pho}
 \end{figure}

\begin{figure}[!htb]
\begin{center}
\resizebox{0.8\columnwidth}{!}{%
    \begin{tabular}{ c  c  c }
   \thead{
   \includegraphics[width=4cm]{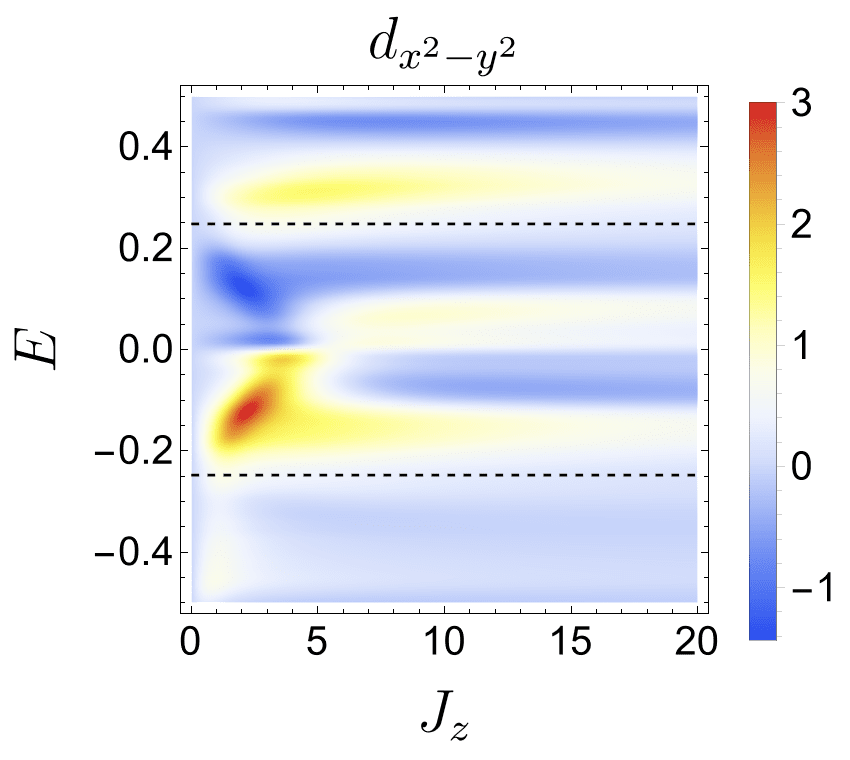}} 
   & \thead{
   \includegraphics[width=4cm]{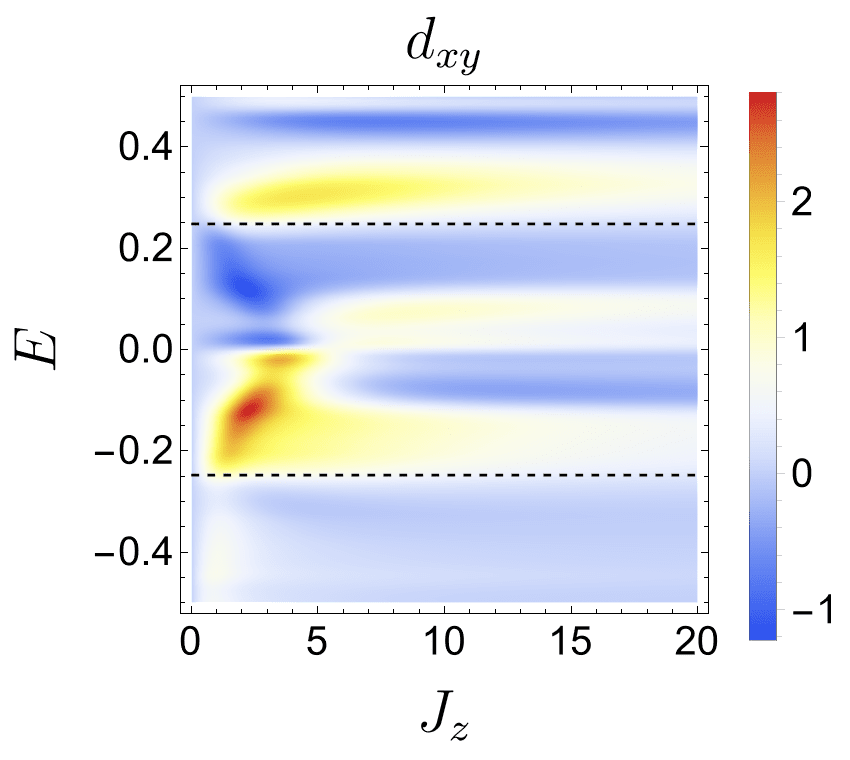}}
       \vspace{-0.4cm}
   \\
    \thead{ 
   \includegraphics[width=4cm]{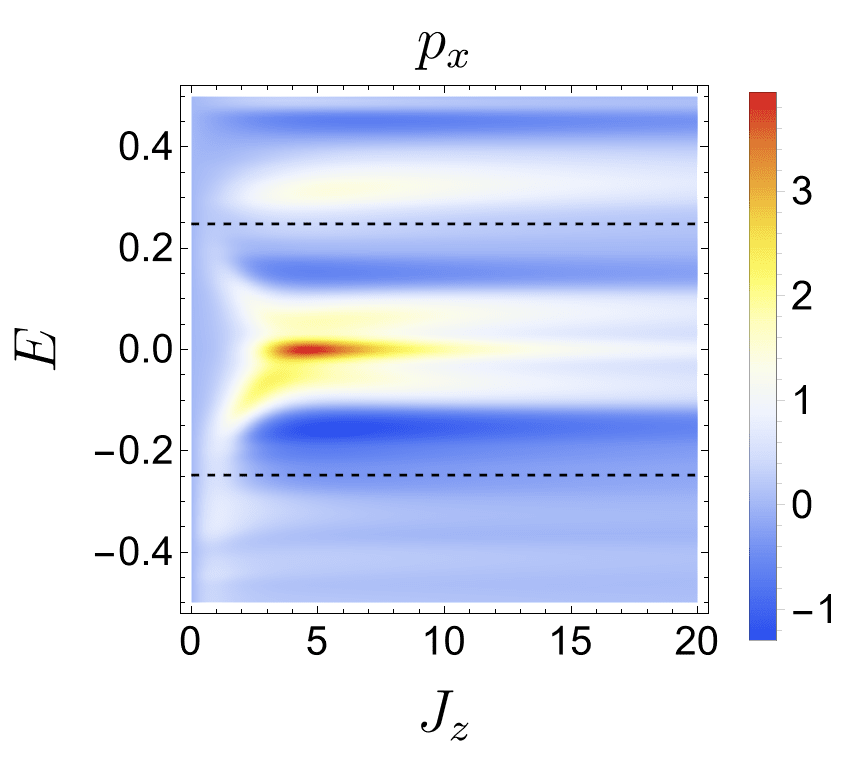}} 
   & \thead{ 
   \includegraphics[width=4cm]{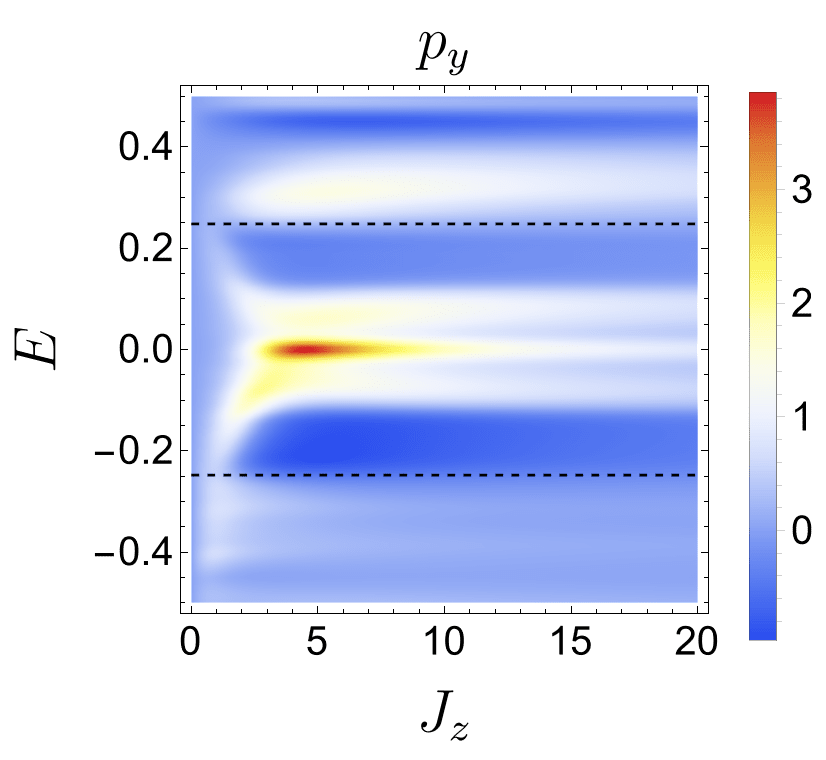}} 
    \end{tabular}
    }%
        \vspace{-0.7cm}
  \end{center}
  \caption{$\delta S_z (E)$ as a function of energy and impurity strength $J_z$ for a $z$-magnetic impurity at $\Delta_0=0.4$ and $\mu = 0.4$ for the nodal SC states in ABC-stacked trilayer graphene. The dotted lines indicate the gap edge.} 
  \label{tril_Shiba_mag_sz}
 \end{figure}

\begin{figure}[!htb]
\begin{center}
\resizebox{0.8\columnwidth}{!}{%
    \begin{tabular}{  c  c c }
     \thead{
   \includegraphics[width=4cm]{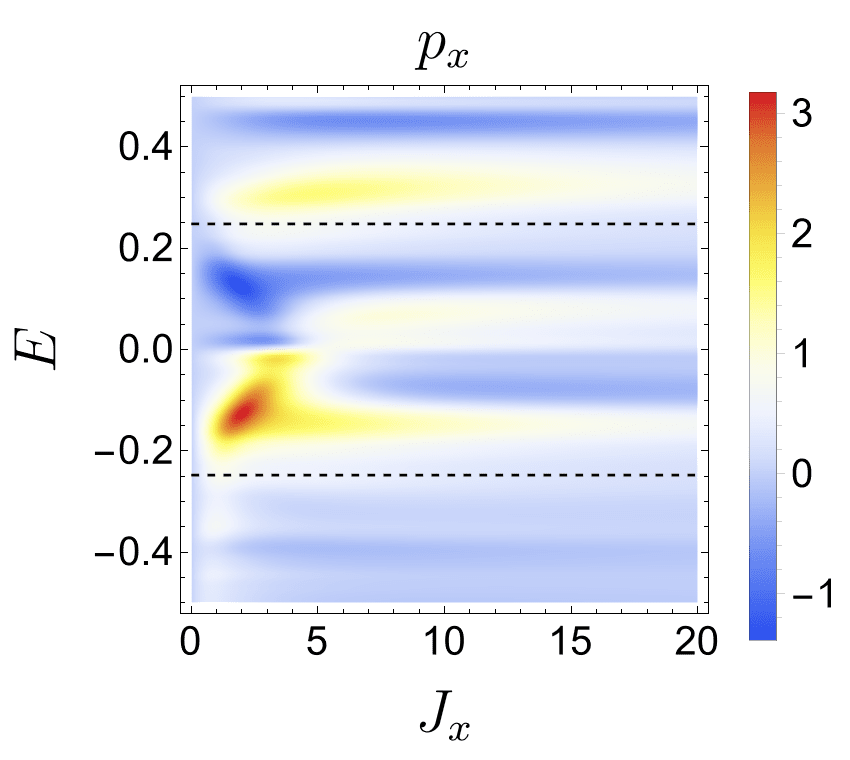}} 
   \thead{ 
   \includegraphics[width=4cm]{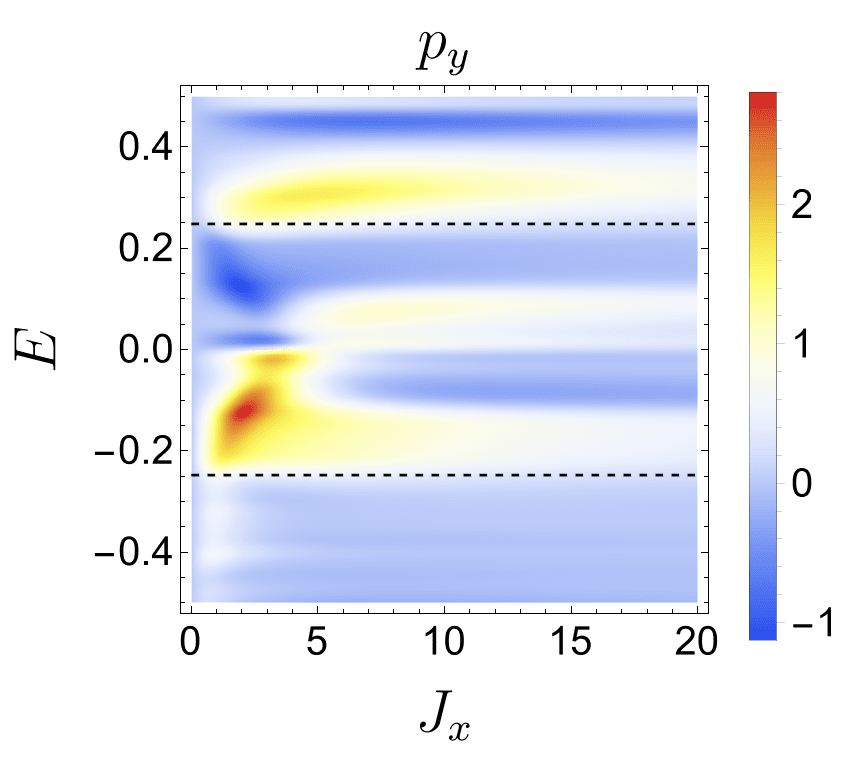}} 
     \end{tabular}
    }%
        \vspace{-0.7cm}
  \end{center}
  \caption{$\delta S_x (E)$  as a function of energy and impurity strength $J_x$ for a $x$-magnetic impurity at $\Delta_0=0.4$ and $\mu = 0.4$ for the spin-triplet nodal SC states in ABC-stacked trilayer graphene. The dotted lines indicate the gap edge.}
  \label{tril_Shiba_mag_sx}
 \end{figure}

Figures~\ref{trilay_Shiba_pho_scal} and \ref{tril_Shiba_mag_pho} show $\delta \rho (E)$ in ABC-stacked trilayer graphene for all nodal order parameters in the presence of a scalar impurity and a $z$-magnetic impurity, respectively. Since in our calculations, the LDOS and SPLDOS are averaged over all atoms in all layers, the results do not depend of the position of the impurity chosen, so we here arbitrary consider an impurity in the top layer, located on the atom that does not sit on top of any atoms in the neighboring layer.  We note that for ABC-stacked trilayer graphene extra subgap states appear beside the two subgap states observed in all the other graphene systems. We have checked using tight-binding calculations that even when these states are close to zero energy for extended parameter ranges, they however do not appear to correspond to Majorana zero modes. Furthermore, Figure~\ref{tril_Shiba_mag_sz} plots $\delta S_z (E)$ for the $z$-magnetic impurity. For completeness we show in Fig.~\ref{tril_Shiba_mag_sx} the effect of changing the spin orientation for the spin-triplet nodal states by plotting $\delta S_x (E)$ for a $x$-magnetic impurity. Same as before, we find that the $x$-magnetic impurity shows a different behavior compared to $y$- and $z$-magnetic impurities for spin-triplet order parameters. 

\begin{figure*}[!htb]
\begin{center}
\resizebox{1.6\columnwidth}{!}{%
    \begin{tabular}{  c  c  c  c}
      \thead{
   \includegraphics[width=6cm]{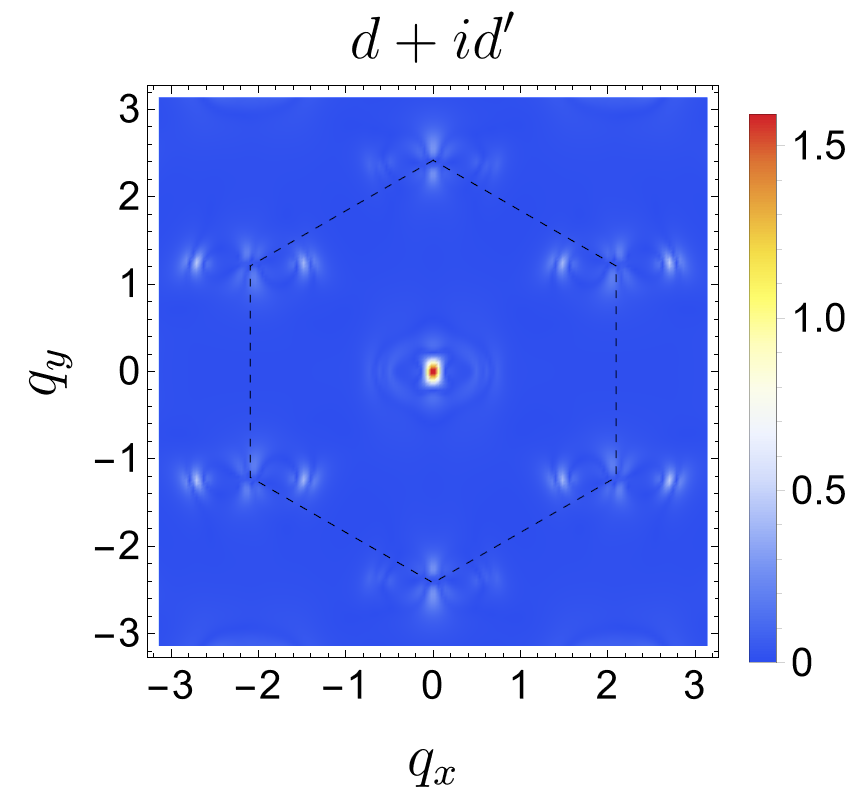}} 
   & \thead{
   \includegraphics[width=6cm]{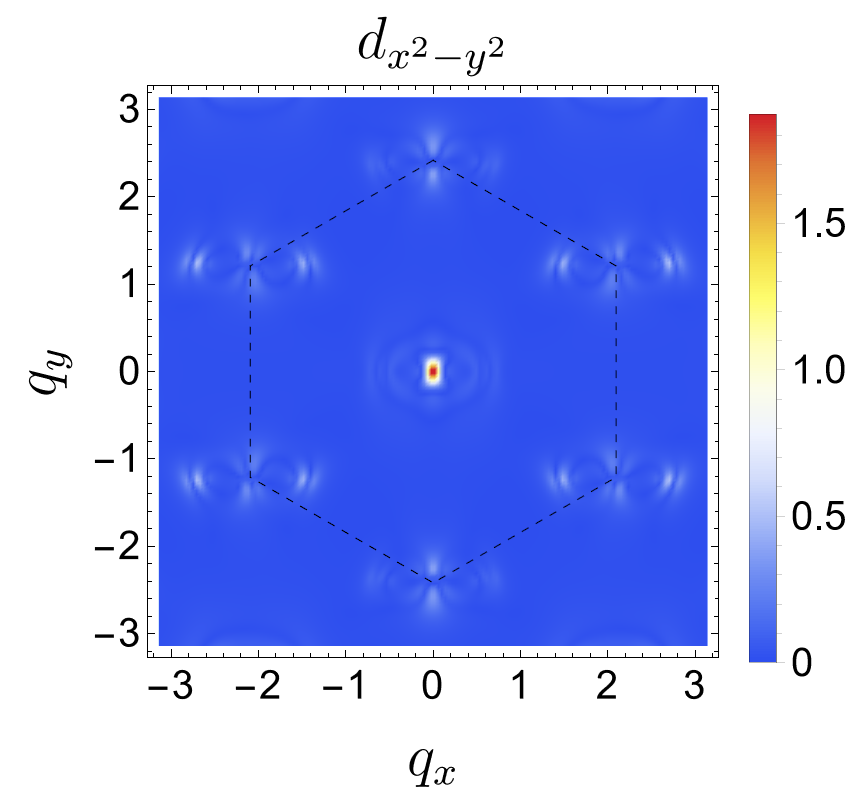}} 
   & \thead{
   \includegraphics[width=6cm]{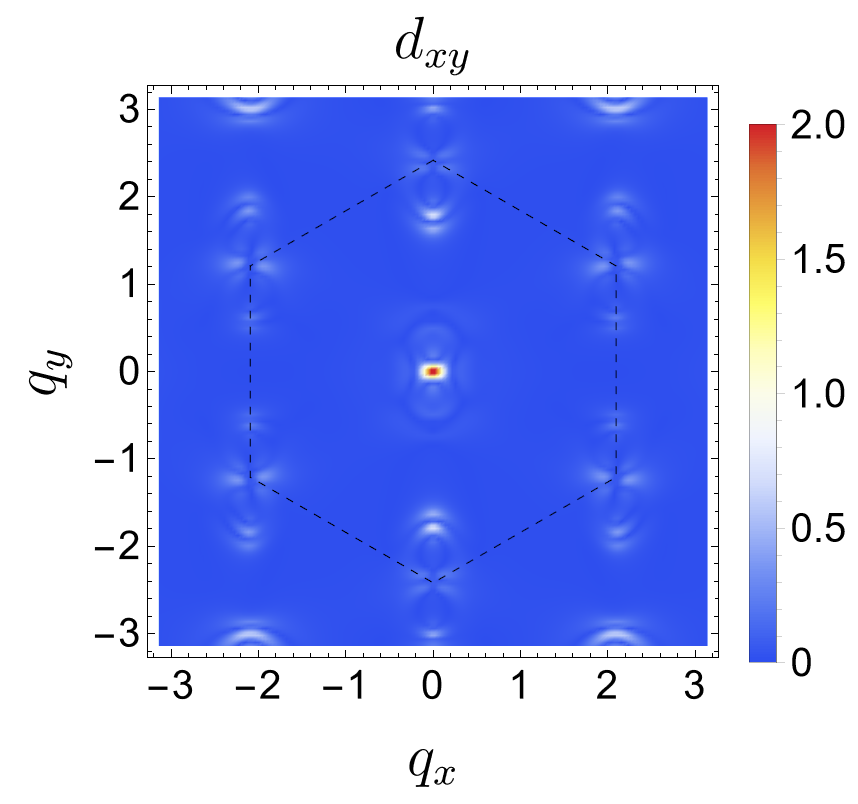}}
       \vspace{-0.4cm}
   \\
      \thead{\includegraphics[width=6cm]{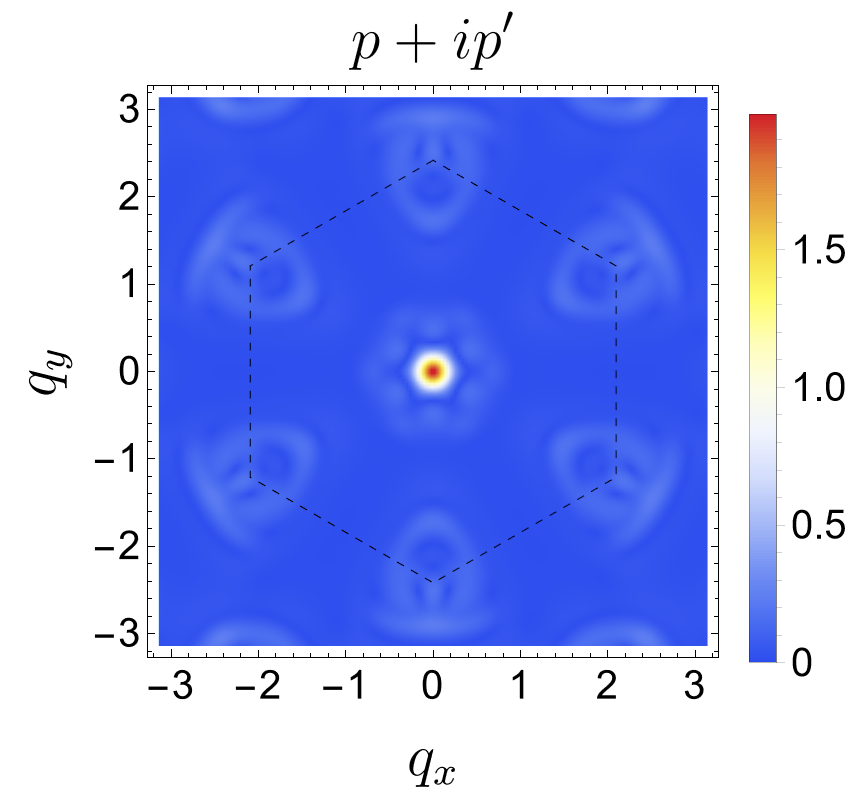} } 
      & \thead{
   \includegraphics[width=6cm]{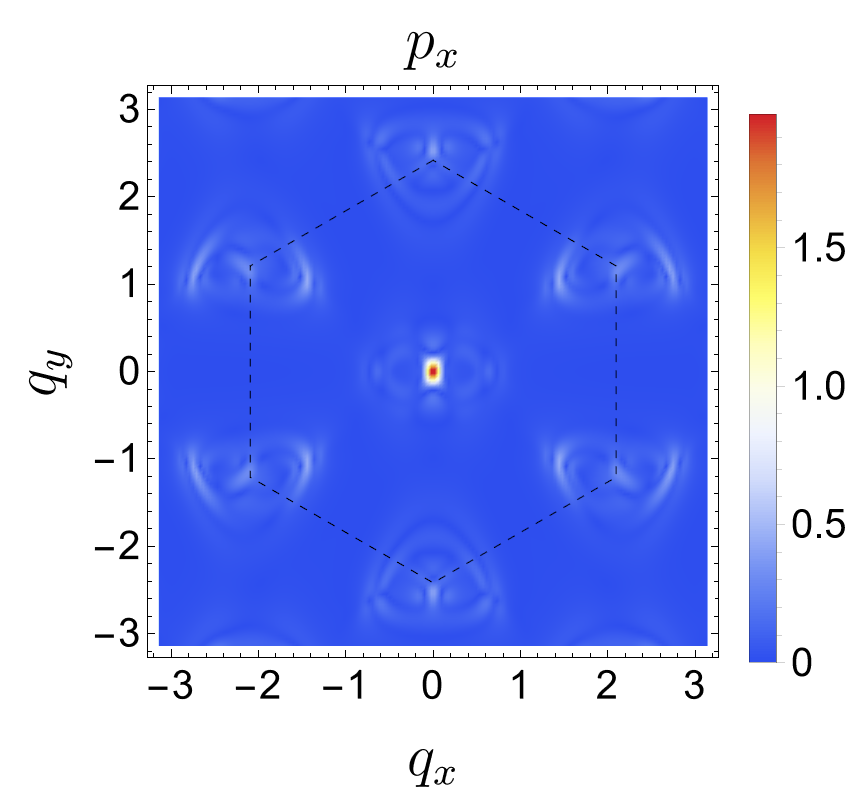}} 
   & \thead{ 
   \includegraphics[width=6cm]{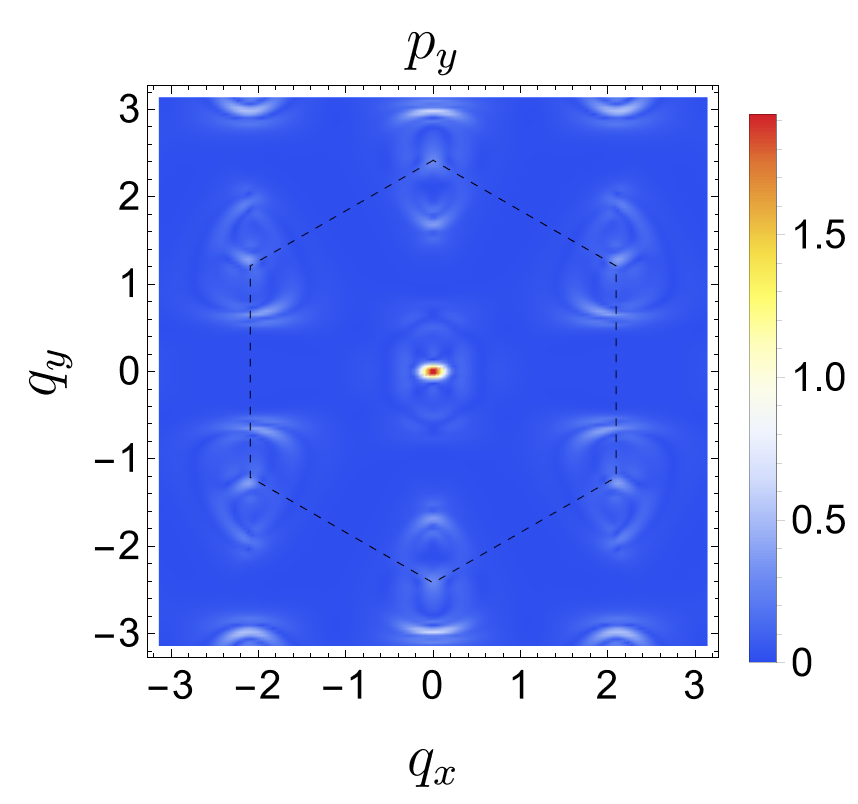}} 

  %\thead{
   %\includegraphics[width=4cm]{b_bottom_pho_ON_g1_02_g3_0.png}} 
   & \thead{
   \includegraphics[width=6cm]{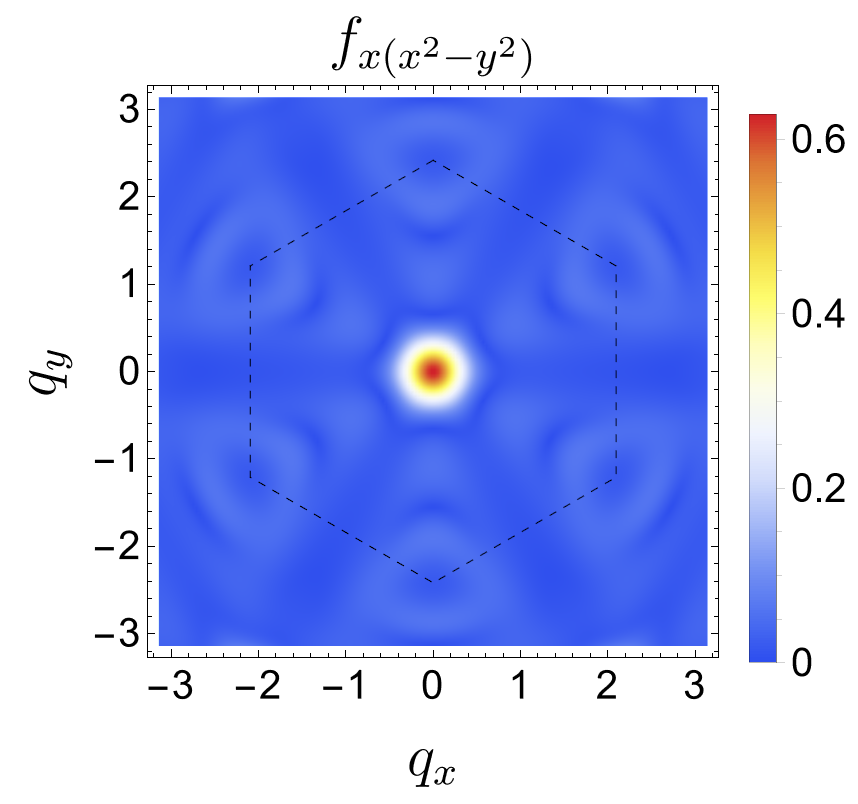}} 
    %\thead{
   %\includegraphics[width=4cm]{b_bottom_pho_s_g1_02_g3_0.png}}
         \end{tabular}
      }%
          \vspace{-0.7cm}
  \end{center}
  \caption{$|\delta \rho (\textbf{q})|$ at zero energy and the corresponding impurity strength values $U$ in Table~\ref{Table3}, evaluated in the top layer for a scalar impurity placed in the bottom layer on an $A$-sublattice atom. We take $\Delta_0=0.4$ and $\mu = 0.4$. The Brillouin zone is indicated by dashed lines.}
  \label{bil_QPI_scal_pho_bottom}
 \end{figure*}

\begin{figure*}[!htb]
\begin{center}
\resizebox{1.6\columnwidth}{!}{%
    \begin{tabular}{  c  c  c  c}
      \thead{
   \includegraphics[width=5cm]{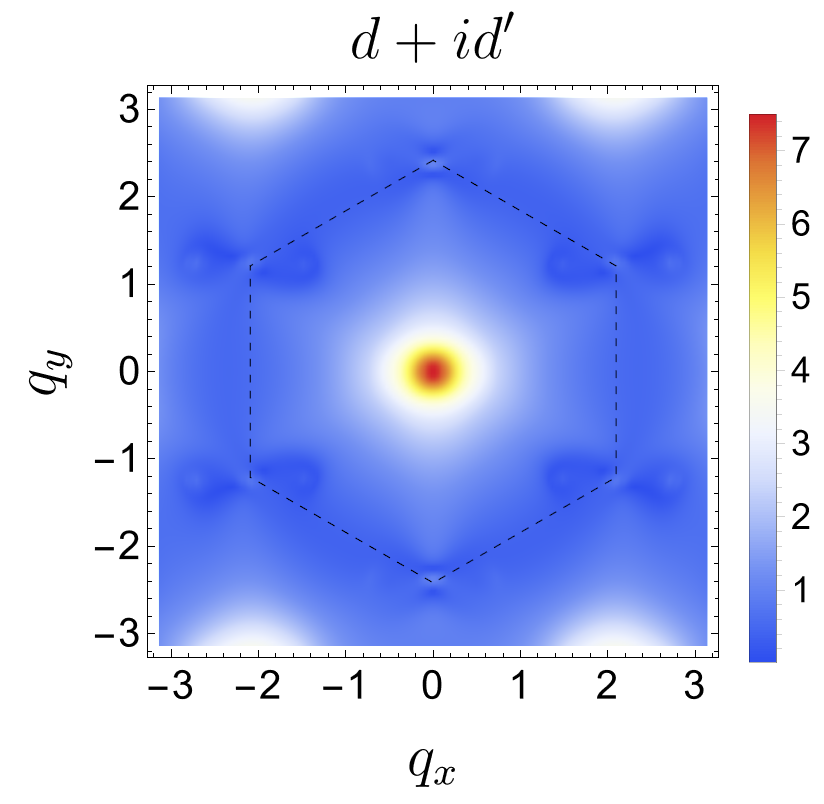}} 
   & \thead{
   \includegraphics[width=5cm]{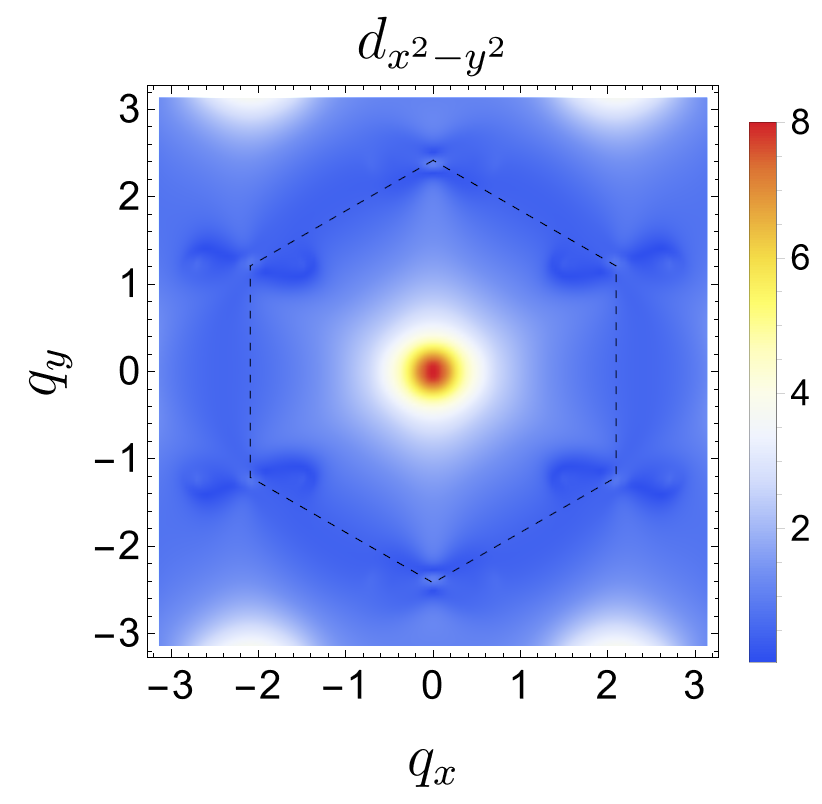}} 
   & \thead{
   \includegraphics[width=5cm]{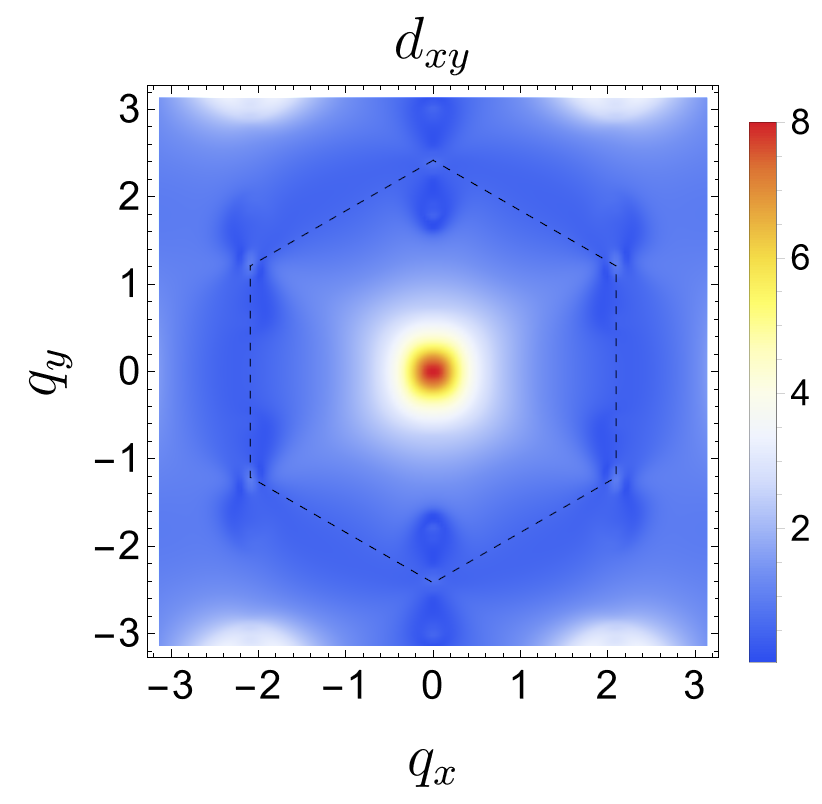}}
       \vspace{-0.4cm}
   \\
      \thead{\includegraphics[width=5cm]{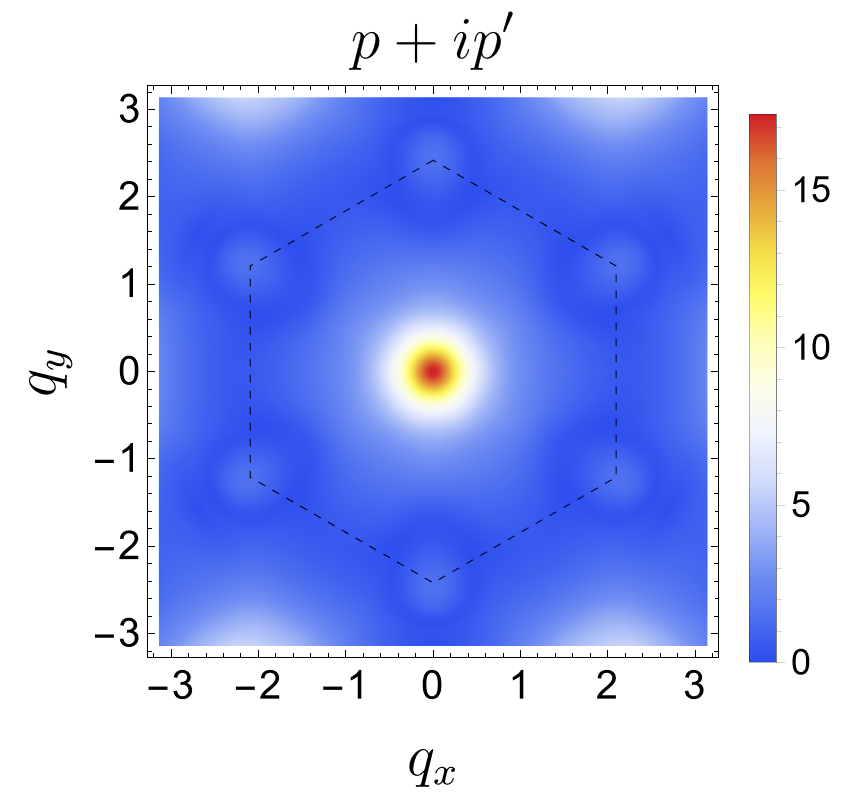} } 
      & \thead{ 
   \includegraphics[width=5cm]{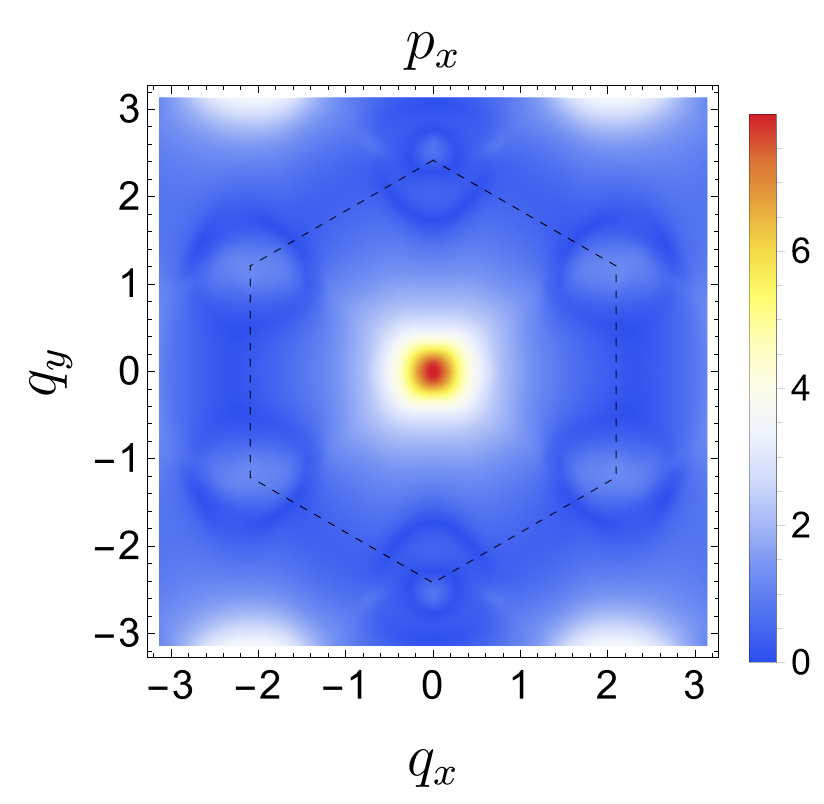}} 
   & \thead{
   \includegraphics[width=5cm]{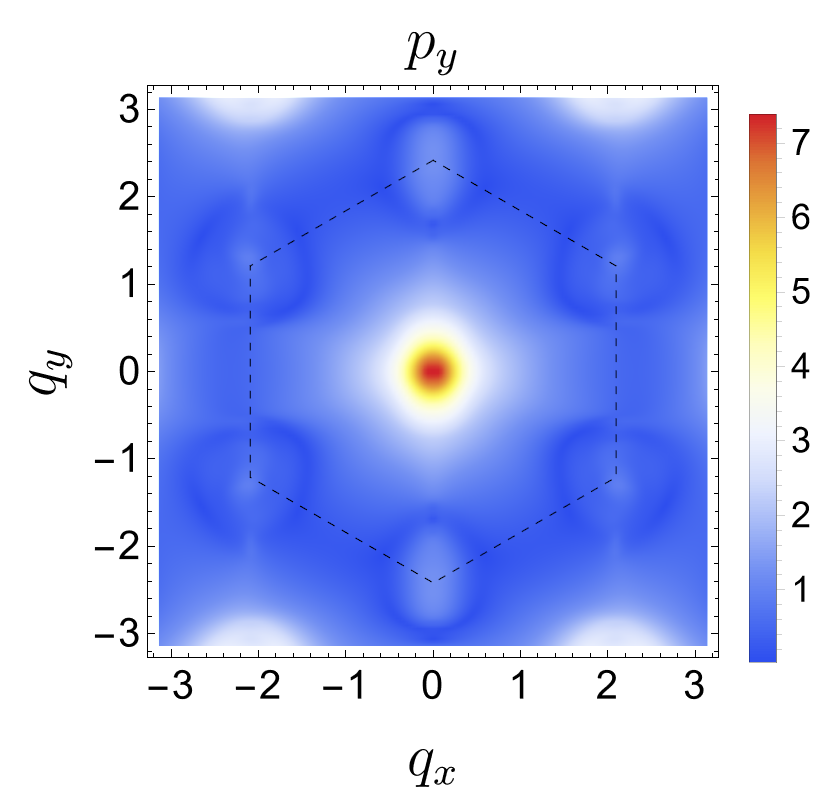}} 

  %\thead{
   %\includegraphics[width=4cm]{b_top_pho_ON_g1_02_g3_0.png}} 
   & \thead{
   \includegraphics[width=5cm]{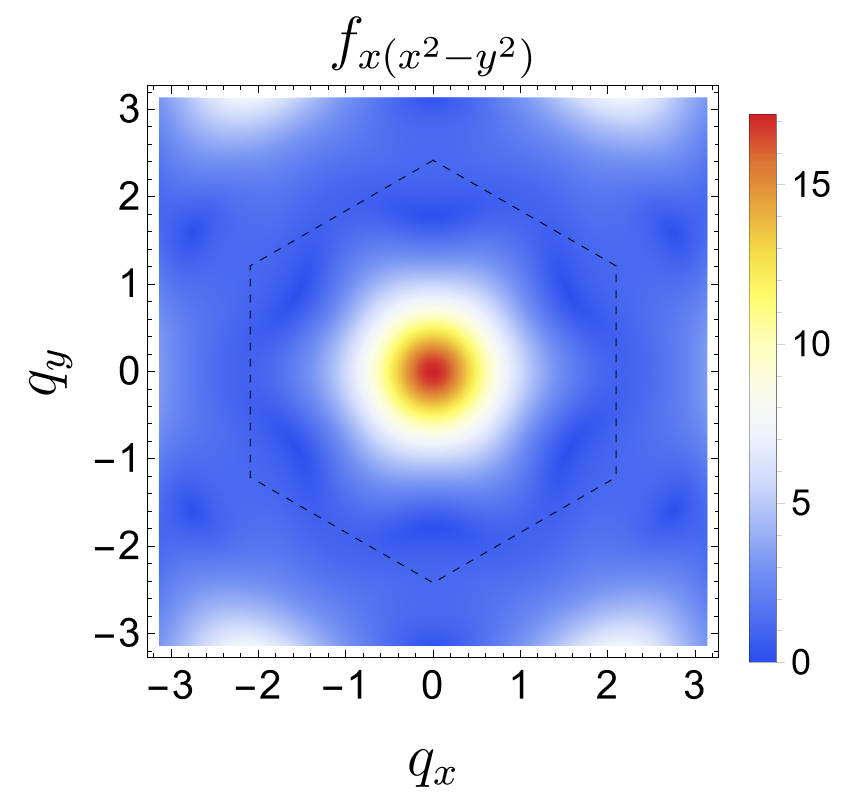}} 
    %\thead{
   %\includegraphics[width=4cm]{b_top_pho_s_g1_02_g3_0.png}}
      \end{tabular}
      }%
          \vspace{-0.7cm}
  \end{center}
  \caption{
  $|\delta \rho (\textbf{q})|$ at zero energy and the corresponding impurity strength values $U$ in Table~\ref{Table3}, evaluated in the top layer for a scalar impurity placed in the top layer on an $A$-sublattice atom. We take $\Delta_0=0.4$ and $\mu = 0.4$. The Brillouin zone is indicated by dashed lines.}
  \label{bil_QPI_scal_pho_top}
 \end{figure*}

We next turn to the QPI patterns. For simplicity we focus first only on the zero energy plots for AB-bilayer graphene in the presence of a scalar impurity. The corresponding impurity strength values are quasi-identical to those for the monolayer and thus we use the same values as those presented in Table~\ref{Table3}. These results can be generalized to the other configurations as we find the differences from the monolayer analysis to be quite generic. We here choose to calculate only the contribution to the LDOS from the top layer atoms, since this is what is measured experimentally \cite{joucken2021direct,kaladzhyan2021quasiparticle1}. Similarly to in Refs.~\onlinecite{joucken2021direct, kaladzhyan2021quasiparticle1} there is a difference in the QPI patterns if the impurity is placed in the top layer or in the bottom layer. However, the measured QPI for a given sample becomes an average between all possible contributions given a random distribution of impurities between the atoms in the two layers.  In Fig.~\ref{bil_QPI_scal_pho_bottom} we plot the QPI resulting from a bottom-layer $A$-sublattice impurity, while in Fig.~\ref{bil_QPI_scal_pho_top} we plot the QPI from a top-layer $A$-sublattice impurity. Note that here in the top layer the $A$ atom is the atom that does not sit on top of another atom, while the $A$ atom in the bottom layer is the one sitting directly underneath another atom. 
We find that the main difference for a bottom-layer impurity, as compared to a top-layer impurity, consists in having a more equal intensity between the central feature at the $\Gamma$ points (corresponding to intra-nodal scattering) and the features at the corners of the Brillouin zone (corresponding to inter-nodal scattering). Thus, for the $A$ bottom-layer impurity the corner features appear sharper. Note also that the features exhibit an extra split due to the interband effect introduced by the interlayer hopping as compared to monolayer graphene.

For ABA- and ABC-stacked trilayer graphene we overall obtain similar QPI patterns (not shown), underwriting the generic features that we observe for all QPI patterns: (i) breaking of the six-fold rotational symmetry for the nodal $d$-wave and $p$-wave states, and more pronounced for the $d$-wave symmetries; (ii) details of the QPI patterns depending on the nature of the impurity (magnetic or scalar), as well as of its spin direction for the spin-triplet states; and (iii) spin-polarized measurements helping to distinguish better between various order parameter symmetries. Overall, these features could help to experimentally identify the symmetry of the SC states in different mono- and few-layer graphene systems.

\section{Conclusion\label{sec:conclusion}}
We calculated the impurity-induced LDOS and SPLDOS, as well as their Fourier transforms (through the QPI patterns), for SC monolayer, AB-stacked bilayer and ABA- and ABC-stacked trilayer graphene, for all expected SC order parameters resulting from NN pairing, or NNN in the case of $f$-wave symmetry. We analyzed the formation of subgap states as a function of energy and impurity strength and found that the number of subgap bound states depends on the type of order parameter. For a scalar impurity we find no subgap states for $s$-wave, both on-site and extended $s$-wave, while two spin-degenerate subgap states appear for all other order parameter symmetries. For a magnetic impurity we find two subgap states for order parameters with $s$-wave symmetries and for nodal states, $d_{xy}$-, $d_{x^2-y_2}$-, $p_x$-, and $p_y$-wave symmetries, while four subgap states exist the fully gapped $d+id^{\,\prime}$-, $p+ip^{\,\prime}$-, and $f$-wave states. The spin polarization of the impurity states is also different if one has a spin-singlet or spin-triplet order parameter and could thus be used to distinguish between the two. In particular the spin-triplet SC states are the only ones for which the opposite-energy subgap states may have an identical spin polarization, and for which the spin structure of the subgap states may depend on the direction of the impurity spin.  These observations could provide an experimental test to distinguish unambiguously via spin-polarized STM between spin-singlet and spin-triplet SC order parameters, as well as between gapped and nodal pairing. The analysis of the QPI patterns additionally show a breaking of the six-fold symmetry for nodal states, while the gapped states preserve this crystalline symmetry, in agreement with the observation that these states have also a symmetry-preserving SC band structure \cite{pangburn2022superconductivity}. Except for a few peculiar situations, our results do not change significantly for bilayer or trilayer graphene, such that we can easily extend our conclusions to multilayer graphene and thus the features described here are quite generic and independent of the number of layers or the graphene layer stacking.

\acknowledgments
ABS acknowledge financial support from the Swedish Research Council (Vetenskapsr\aa det Grant No.~2018-03488) and the Knut and Alice Wallenberg Foundation through the Wallenberg Academy Fellows program. NS would like to thank the National Science Centre (NCN, Poland) for funding under the grant 2018/29/B/ST3/01892.

\end{document}